\documentclass[5p,times]{elsarticle}

\usepackage[numbers]{natbib}
\usepackage{float}
\usepackage{subfig}
\usepackage{graphicx}
\usepackage{url}
\usepackage{xcolor}
\usepackage{stfloats}
\usepackage{tcolorbox}
\usepackage{amsmath}
\usepackage{amsfonts}
\usepackage{multirow}
\usepackage{booktabs}
\usepackage{listings}
\usepackage{pifont}
\usepackage{makecell}
\usepackage[ruled,vlined]{algorithm2e} 
\usepackage{amsmath,amssymb}           
\usepackage{amsthm}
\usepackage{graphicx}
\usepackage{subfig}

\theoremstyle{definition}
\newtheorem{definition}{Definition}

\usepackage{nicematrix}
\usepackage{bm}
\usepackage{orcidlink}
\usepackage{threeparttable}

\definecolor{verylightgray}{rgb}{.97,.97,.97}

\newcommand{\tool}{ARIADNE}

\journal{}

\begin{document}

\begin{frontmatter}
	
	\title{{\tool}: Agentic Reward-Informed Adaptive Decision Exploration via Blackboard-Driven MCTS for Competitive Program Generation}

\author[NTU]{Minnan Wei\orcidlink{0009-0007-5479-5784}\corref{mycorrespondingauthor}}
\ead{minnanvvei@gmail.com}	

\author[NTU]{Xiang Chen\orcidlink{0000-0002-1180-3891}\corref{mycorrespondingauthor}}
\cortext[mycorrespondingauthor]{Corresponding authors}
\ead{xchencs@ntu.edu.cn}	

\author[NTU]{Xiaoshuai Niu\orcidlink{0009-0009-1434-3550}}
\ead{2145290096@qq.com}

\author[NTU]{Siyu Chen\orcidlink{0009-0006-8951-4922}}
\ead{chensiyu043@gmail.com}
		
\address[NTU]{School of Artificial Intelligence and Computer Science, Nantong University, Nantong, China}

\begin{abstract}
Competitive program generation aims to automatically produce correct and efficient solutions for programming-contest problems under strict time and memory constraints. Existing LLM-based approaches often fail to perform explicit algorithmic planning and to handle edge cases robustly, leading to unreliable one-shot generation. Moreover, although execution feedback is essential for iterative debugging and refinement, incorporating such feedback effectively within limited computational budgets remains difficult. To overcome these limitations, we propose {\tool}, a blackboard-driven Monte Carlo Tree Search (MCTS) framework that models program generation as a sequential decision process. {\tool} organizes the generation workflow into five coordinated stages (i.e., strategy selection, code generation, test generation, quality evaluation, and code repair) while maintaining a shared blackboard that accumulates structured evidence to guide subsequent decisions. Experiments on four benchmarks (APPS, CodeContests, CodeContests+, and LiveCodeBench) show that {\tool} consistently achieves the best Pass@1 performance across multiple LLM backends. With GPT-4o, {\tool} attains Pass@1 scores of 41.30, 46.67, 27.27, and 20.91, surpassing the strongest baseline CodeSim by up to 26.06 points, while further improvements are observed with DeepSeek-V3.2. These results indicate that combining global search through MCTS with persistent evidence accumulation on a shared blackboard enables systematic exploration and effective feedback utilization, substantially enhancing the capability of LLMs in competitive program generation.

\end{abstract}


\end{frontmatter}

\section{Introduction}
\label{sec:intro}

Competitive program generation targets the automatic synthesis of correct and efficient solutions to programming-contest problems under strict time/memory constraints and hidden test suites, making it a compelling benchmark for algorithmic reasoning and practical code reliability. Despite strong general coding ability, Large Language Models (LLMs) still struggle in this setting because success often hinges on constructing accurate problem-solving strategies, including multi-step logical reasoning, explicit algorithmic planning, and comprehensive edge-case handling, that go beyond basic functional code generation~\cite{hossain2025llm, ouyang2025empirical, wang2024large, esfahani2024understanding, abbassi2025unveiling, dinh2023large, liu2024exploring}.  

Prior work~\cite{islam2024mapcoder,islam2025codesim,pan2025codecor,xu2024sra,chen2025treemind,delorenzo2024make,han2025exploring,salemi2025llm} has explored several directions to improve LLMs on competitive program generation, including iterative agent-style refinement, search-based exploration, and shared-workspace designs that externalize intermediate artifacts for better context management. 
However, when used in isolation, these paradigms still leave critical gaps under contest constraints: agentic pipelines are often driven by pre-fixed workflows that limit adaptive decision-making; MCTS-based generation frequently lacks persistent and structured evidence reuse across branches; and blackboard systems typically do not incorporate an explicit global planner to allocate search budget effectively and recover from early suboptimal decisions.

Existing paradigms leave the following practical gaps under contest constraints: (1) agentic pipelines often rely on pre-specified workflows that are brittle when early assumptions are wrong; (2) search-based generation may explore broadly but fails to accumulate and reuse structured evidence across branches; and (3) blackboard-style coordination stores artifacts, yet typically lacks an explicit mechanism to prioritize actions and \emph{reallocate} budget based on what has been learned so far.  

To address these gaps, we propose {\tool}
(\textbf{A}gentic \textbf{R}eward-\textbf{I}nformed \textbf{A}daptive \textbf{D}ecisio\textbf{N} \textbf{E}xploration), a blackboard-driven MCTS framework for competitive program generation. {\tool} views solution construction as iterative decision making over actions (i.e., strategy selection, code generation, test generation, quality evaluation, and code repair) conditioned on a shared blackboard state.  Specifically, \emph{Reward-Informed} means that backpropagated rewards do not merely rank complete candidates, but actively steer exploration toward decision points that are most likely to yield correctness and efficiency under a limited budget. Meanwhile, \emph{Adaptive Decision} means that structured diagnostics are written back to the blackboard and reused as persistent state, so subsequent expansions are conditioned on accumulated evidence rather than isolated rollouts. As a result, {\tool} can (a) deviate from rigid agent workflows when new evidence contradicts earlier plans, (b) reuse failure evidence to avoid repeating unproductive branches, and (c) dynamically concentrate search on the most informative fixes and promising strategy alternatives.

We first evaluate {\tool} on four representative competitive-programming benchmarks (APPS~\cite{hendrycks2021measuring}, CodeContests~\cite{li2022competition}, CodeContests+~\cite{wang2025codecontests+}, and LiveCodeBench~\cite{jain2024livecodebench}), which vary in problem domains, difficulty, and evaluation protocols. We compare against three state-of-the-art agentic baselines, including MapCoder~\cite{islam2024mapcoder}, CodeSim~\cite{islam2025codesim}, and CodeTree~\cite{li2025codetree}, and report Pass@1 together with runtime statistics. Across all benchmarks, {\tool} consistently achieves the best Pass@1 in matched settings. For example, with GPT-4o, it reaches 41.3\% in the APPS benchmark and reaches 46.67\%, 27.27\%, and 20.91\% in CodeContests, CodeContests+, and LiveCodeBench benchmarks, respectively.

We then assess real contest performance using complete ICPC/CCPC-style multi-problem contest instances collected from Codeforces, including the 2025 ICPC Asia Shenyang Regional Contest and the 2025 CCPC Fujian Invitational. Under the same contest protocol and submission budgets, {\tool} improves pass@k behavior and solved problem numbers over the best baseline from the benchmark comparison; for instance, on Shenyang Regional Contest, it attains pass@1/3/5 of 3/13, 6/13, and 7/13 (vs.\ 1/13, 4/13, and 5/13), and on Fujian Regional Contest, it reaches 4/13, 5/13, and 7/13 (vs.\ 2/13, 2/13, and 5/13). 

Our results indicate that competitive program generation benefits from explicit global planning over heterogeneous decisions (strategy selection, code synthesis, testing, and repair) rather than fixed agent pipelines.
By coupling a persistent blackboard state with reward-informed MCTS,  {\tool} can accumulate and reuse diagnostic evidence to steer exploration toward high-value solution trajectories.

The main contributions of our study can be summarized as follows:
\begin{itemize}
  \item We propose {\tool}, 
  a blackboard-driven MCTS framework that performs reward-informed, adaptive decision exploration for competitive program generation, coordinating strategy selection, code generation, quality evaluation, and code repair within a unified search process.
  \item We design a structured blackboard system and action space that enable evidence accumulation and reuse, allowing search to adapt its decisions based on execution feedback.
  \item We conduct extensive evaluations on diverse benchmarks and realistic recent ICPC Regional Contests, demonstrating consistent improvements over strong baselines.
\end{itemize}

\noindent{\textbf{Open Science}} To enable the replication of our research, we share our dataset, source code, and detailed results on GitHub (\url{https://github.com/minnanWei/ARIADNE}).

\noindent{\textbf{Paper organization.}} 
The remainder of this paper is organized as follows: Section~\ref{sec:rbrm} introduces the background and motivation for reward-informed agentic search in competitive-program generation. Section~\ref{sec:method} presents {\tool}, detailing the blackboard-driven framework, reward modeling, and the adaptive decision exploration procedure based on MCTS. Section~\ref{sec:setup} describes the experimental setup, including research questions, datasets, baselines, metrics, and implementation details. Section~\ref{sec:results} reports the experimental results and key findings on benchmark tasks and full-contest evaluations. Section~\ref{sec:analysis} further discusses generalization across different LLM backbones, efficiency statistics (e.g., token usage and execution cost), and threats to validity. Section~\ref{sec:related} reviews related work and highlights the novelty of our study. Section~\ref{sec:conclusion} concludes the paper and outlines future research directions.

\section{Research Background and Research Motivation}
\label{sec:rbrm}

\subsection{Competitive Program Generation}

Competitive program generation seeks to automatically produce correct and efficient solutions for algorithmic problems drawn from programming contests. Unlike general-purpose code generation, these tasks demand explicit algorithmic strategy selection, multi-step reasoning, strict time and memory constraints, and thorough handling of edge cases.

To support rigorous evaluation, recent benchmarks emphasize both algorithmic diversity and execution-level correctness. ProBench~\cite{yang2025probench} compiles real contest problems annotated with algorithm tags and difficulty levels. TACO~\cite{li2023taco} provides large-scale datasets with fine-grained topic and skill annotations. CodeContests+~\cite{wang2025codecontests+} enhances evaluation fidelity through high-quality test-case generation, while COMPASS~\cite{meaden2025compass} introduces multi-dimensional metrics encompassing correctness, efficiency, and code quality.

\subsection{Monte Carlo Tree Search for Code Generation}

MCTS (Monte Carlo Tree Search) is a search algorithm for sequential decision-making under uncertainty, which balances exploration and exploitation through iterative selection, expansion, simulation, and backpropagation. Its principled management of this trade-off makes MCTS particularly well-suited for reasoning over large combinatorial decision spaces.

Recent work has explored integrating MCTS with LLMs to enhance reasoning and code generation. TreeMind~\cite{chen2025treemind} applies MCTS to explore UI interaction sequences for bug reproduction, while VeriGen+MCTS ~\cite{delorenzo2024make} leverages MCTS to guide RTL code generation under functional and PPA constraints. These studies illustrate that MCTS can serve as an effective global planner for complex software engineering tasks.

\subsection{Blackboard System}

Blackboard systems provide a coordination architecture for multi-agent and multi-knowledge-source systems, in which agents interact indirectly through a shared, structured workspace rather than via direct communication. Rather than functioning as simple shared memory, a blackboard incrementally accumulates intermediate hypotheses, constraints, and partial results, making them available to all agents as a common foundation for subsequent reasoning and action.

This paradigm has been widely adopted to facilitate collaboration among heterogeneous agents in complex decision-making settings. 
Early work~\cite{ito2000blackboard} employed a blackboard-based open-tender mechanism for collaborative negotiation, where multiple agents submit competing proposals and constraints to a shared workspace. In robotics and control systems, blackboard architectures have been used to integrate perception, planning, and execution agents around a shared world model, enabling coordinated behavior without tight coupling between modules. More recently, blackboard systems have been revisited in the context of LLM-based multi-agent systems. Han et al.~\cite{han2025exploring} demonstrate that externalizing intermediate reasoning artifacts to a blackboard mitigates information fragmentation in multi-agent LLM collaboration and supports dynamic selection of subsequent actions. Similarly, Salemi et al.~\cite{salemi2025llm} show that blackboard-mediated coordination allows agents to self-select and contribute based on the evolving shared state, enhancing both flexibility and scalability.

\subsection{Research Motivation}

By analyzing the previous studies~\cite{yang2025probench,li2023taco,wang2025codecontests+,meaden2025compass,chen2025treemind,delorenzo2024make,ito2000blackboard,han2025exploring,salemi2025llm}, we identify the following three key limitations.

\textbf{Limitation 1: Insufficiency of one-shot prompting.} 
Studies such as Wei et al.~\cite{wei2025evaluating} and LLM-ProS~\cite{hossain2025llm} indicate that only a limited portion of contest problems can be solved using basic prompting. A key reason is that competitive program generation is not a simple mapping from problem description to code implementation. Instead, it requires (1) committing to an algorithmic strategy under strict complexity constraints, (2) translating that strategy into a correct implementation, and (3) iteratively correcting subtle failures revealed through execution. Without an explicit mechanism for hypothesis revision and test-driven refinement, one-shot generation tends to overfit to superficial patterns in the problem statement and fails to systematically recover from early incorrect decisions.

\textbf{Limitation 2: Single-trajectory nature of MCTS-based generation.} 
Existing approaches~\cite{xu2024sra,chen2025treemind,delorenzo2024make} that apply MCTS to code generation typically operate along a single reasoning process or generation trajectory. Therefore, the search tends to explore variations of a dominant draft or a single chain of intermediate reasoning, rather than coordinating specialized capabilities (such as strategy analysis, targeted test synthesis, and evidence-driven repair). In addition, intermediate artifacts are often treated as temporary outputs rather than a persistent state, which limits cross-branch knowledge transfer and diminishes exploration efficiency in large program spaces where failures frequently exhibit recurring patterns.

\textbf{Limitation 3: Absence of an explicit global planner in blackboard coordination.} 
While blackboard systems facilitate information sharing and coordination among agents~\cite{ito2000blackboard,han2025exploring,salemi2025llm}, the evolution of the shared workspace is often governed by fixed workflows, local heuristics, or reactive triggering rules. In such settings, agents may contribute useful artifacts, but there is no principled mechanism for allocating search budget across competing strategy hypotheses, balancing exploration and exploitation, or systematically revisiting earlier decisions in light of new evidence. As a result, the system can become path-dependent: early suboptimal strategy commitments or premature repairs may dominate subsequent steps, constraining the exploration of alternative solution directions and reducing robustness on challenging contest problems.

\section{Our Proposed Framework {\tool}}
\label{sec:method}

In this section, we present {\tool}, a blackboard-driven MCTS framework for competitive program generation. The primary goal of {\tool} is to systematically explore the solution space and progressively converge to an optimal program for a given algorithmic problem, by coupling search-based planning with modular agent execution and a blackboard system that maintains structured shared knowledge across steps and branches. Rather than treating code generation as a one-shot synthesis problem, {\tool} models competitive programming as a sequential decision-making process, where partial solutions are iteratively generated, evaluated, repaired, and refined under explicit correctness constraints. The framework comprises three key components:

\begin{itemize}
\item \textbf{MCTS}: which functions as a global planner that selects among competitive program drafts via a UCB-guided tree policy, balancing exploration and exploitation, and shifting the search toward branches with higher estimated value as the node statistics $p(u)$ are updated through backpropagation;

\item \textbf{Modular agents}, which execute concrete actions such as code generation, quality evaluation, test generation, and code repair, thereby enabling diverse solution attempts beyond a single draft;

\item \textbf{A blackboard system}, which maintains a structured shared workspace for intermediate constraints, tests, and other artifacts, enabling information to persist and be systematically reused throughout the search process.
\end{itemize}

\begin{figure*}[htbp] 
    \centering \includegraphics[width=0.85\textwidth]{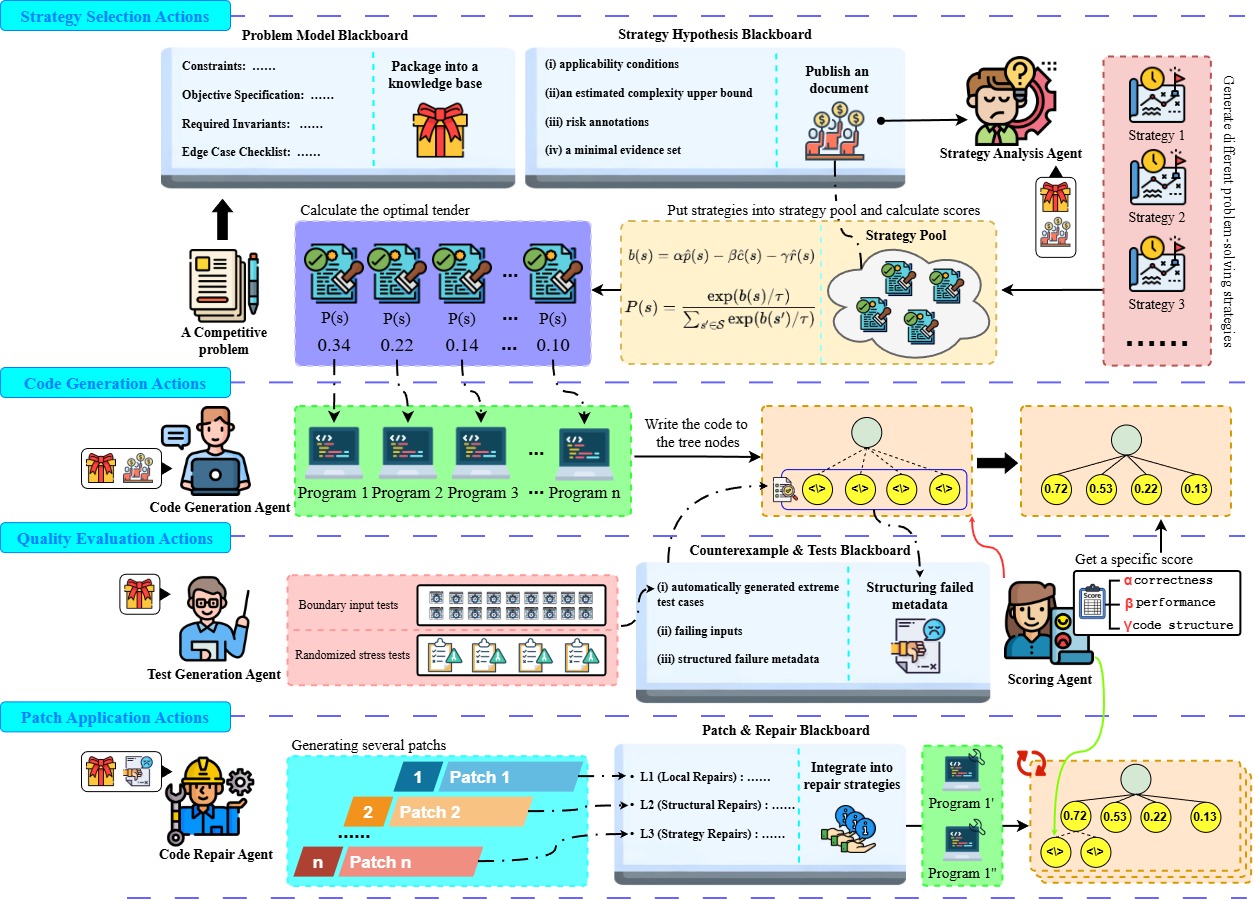}
\caption{Overview of {\tool}: Illustration of all state-transition actions, including Strategy Selection, Code Generation, Evaluation, and Patch Application.} 
    \label{Fig:ARIADNE} 
\end{figure*}

Figure~\ref{Fig:ARIADNE} illustrates the overall architecture of {\tool}. Given a competitive programming problem, the system initializes a set of blackboard, constructs an MCTS search tree, and iteratively improves program drafts through agent-driven actions guided by MCTS. At each iteration, agents read from and write to the blackboard, allowing quality evaluation feedback to accumulate and inform subsequent expansions. This coupling turns refinement and repair into first-class transitions in the search process, while the global planner prioritizes promising directions without committing irrevocably to early choices.

\subsection{Problem Formulation as Search}
\label{sec:Problem Formulation as Search}

In this study, we model competitive program generation as a tree search problem over the space of program drafts augmented with shared intermediate artifacts. We next introduce the relevant definitions.

\begin{definition}[Problem Instance]
A problem instance is $x$, including the statement, I/O specification, constraints, and sample tests.
\end{definition}

\begin{definition}[State]
A state at step $t$ is $s_t=(c_t,\mathcal{B}_t)$, where $c_t$ is the current program draft and $\mathcal{B}_t$ is a blackboard that stores intermediate artifacts (e.g., parsed constraints, solution hypotheses, counterexamples, and repair notes).
\end{definition}

\begin{definition}[Action]
An action $a_t \in \mathcal{A}(s_t)$ specifies a state transition applied to the current state $s_t=(c_t,\mathcal{B}_t)$ and induces a transition $s_{t+1}=T(s_t,a_t)$.
We define the action space as a union of five action families:
\begin{equation}
\mathcal{A}(s_t)=\mathcal{A}_{\text{str}}(s_t)\ \cup\ \mathcal{A}_{\text{gen}}(s_t)\ \cup\ \mathcal{A}_{\text{test}}(s_t)\ \cup\ \mathcal{A}_{\text{eval}}(s_t)\ \cup\ \mathcal{A}_{\text{patch}}(s_t).
\end{equation}
where
\begin{itemize}
    \item \textbf{Strategy Selection Actions} ($\mathcal{A}_{\text{str}}$): activate, rank, or refine algorithmic strategy hypotheses stored on the blackboard.
    \item \textbf{Code Generation Actions} ($\mathcal{A}_{\text{gen}}$): synthesize or revise candidate implementations under the selected strategy and blackboard context, producing an updated draft $c_{t+1}$.
    \item \textbf{Test Generation Actions} ($\mathcal{A}_{\text{test}}$): construct discriminative tests and counterexamples to improve screening and failure localization, updating the shared test/evidence entries in $\mathcal{B}$.
    \item \textbf{Quality Evaluation Actions} ($\mathcal{A}_{\text{eval}}$): assess candidate drafts through staged testing/analysis, producing a scalar reward and structured diagnostics that update $\mathcal{B}$.
    \item \textbf{Patch Application Actions} ($\mathcal{A}_{\text{patch}}$): apply structured repairs or targeted modifications to obtain a patched program $c_{t+1}$, together with repair-related blackboard updates.
\end{itemize}
Each action is instantiated by invoking its corresponding agents.
\end{definition}

\begin{definition}[Transition]
Executing $a_t$ produces a successor node $s_{t+1}=T(s_t,a_t)=(c_{t+1},\mathcal{B}_{t+1})$, where the program draft and blackboard are updated according to the action outcome.
\end{definition}

\begin{definition}[Value]
We define a value function $V(s_t)$ that quantifies the quality of the current draft, primarily based on correctness, and optionally incorporating efficiency and code quality metrics. For non-terminal drafts, we compute a scalar reward $R$ as a weighted combination of these signals:
\begin{equation}
    R = \alpha R_{\text{corr}} + \beta R_{\text{perf}} + \gamma R_{\text{struct}}
\end{equation}
where $R_{\text{corr}}$ is a correctness-related signal (dominated by test outcomes), and $R_{\text{perf}}$ and $R_{\text{struct}}$ measure efficiency and code quality, respectively. In our experiments, we set $\alpha$ = 0.6 and $\beta$ = $\gamma$ = 0.2\footnote{We assign a larger weight to $\alpha$ to prioritize functional correctness in the reward. In particular, the test-based correctness signal serves as the primary gate: a draft that passes the evaluation tests is treated as ``passed", whereas a draft that fails any test is treated as ``failed".}. The evaluation additionally produces structured feedback, such as failing test cases and repair hints, which is recorded in $\mathcal{B}$ to guide subsequent actions.
\end{definition}

Given the above definitions, program generation can be formulated as a sequential decision process. Starting from an initial state $s_0 = (c_0, \mathcal{B}_0)$,
consisting of an empty code draft and an initialized blackboard, the algorithm repeatedly selects an action $a_t$ from the action space $A(s_t)$. It applies it to obtain a refined draft $c_{t+1}$ and updated artifacts $\mathcal{B}_{t+1}$ on related blackboards. The goal is to identify a final solution program $p_{\text{sol}}$ that is functionally correct and, ideally, efficient within a bounded computational budget. This formulation naturally motivates the use of tree search methods to balance exploration of diverse solution hypotheses with exploitation of high-value partial drafts.

\begin{definition}[Terminal Condition]
A node is considered terminal if the current program passes all tests under the evaluation protocol or if the search budget is exhausted. In either case, the final draft $c_m$ is treated as the solution program for the problem and is denoted by $p_{\text{sol}}$.
\end{definition}

\begin{algorithm}[!htbp]
\caption{Blackboard-Driven MCTS for Competitive Program Generation}
\label{alg:ariadne_mcts}
\LinesNumbered
\BlankLine
\KwIn{problem instance $x$; iteration budget $K_{\max}$; UCB coefficient $C$; temperature $\tau$; reward weights $(\alpha,\beta,\gamma)$}
\KwOut{solution program $p_{\mathrm{sol}}$ if found; otherwise best draft $c^{\mathrm{best}}$}
\textbf{Initialize root state.}\\
$c_0 \leftarrow \varnothing$\;
$\mathcal{B}_{PM}\leftarrow BuildProblemModel(x)$\;
$\mathcal{B}_{SH}\leftarrow \varnothing\; \mathcal{B}_{CT}\leftarrow \varnothing\; \mathcal{B}_{PR}\leftarrow \varnothing$\;
$\mathcal{B}_0 \leftarrow (\mathcal{B}_{PM},\mathcal{B}_{SH},\mathcal{B}_{CT},\mathcal{B}_{PR})$\;
$s_0 \leftarrow (c_0,\mathcal{B}_0)$\;
$v_0 \leftarrow NewNode(s_0)$\;
$c^{\mathrm{best}}\leftarrow c_0;\; R^{\mathrm{best}}\leftarrow -\infty$\;
\tcp{Each node $x$ maintains $N(x)$ and $\bar{X}(x)$.}

\BlankLine
\For{$k \leftarrow 1$ \KwTo $K_{\max}$}{

    $v \leftarrow v_0$\; $\pi \leftarrow \langle v_0\rangle$\;
    \While{$v$ is fully expanded \textbf{and} non-terminal}{
        \tcp{For child $u\in \mathrm{Child}(v)$: $\displaystyle UCB(u)=\bar{X}(u)+C\sqrt{\frac{\ln(N(v))+1}{N(u)}}$}
        \tcp{Pick $u^\star=\arg\max_{u\in \mathrm{Child}(v)} UCB(u)$, or sample $u^\star$ via softmax over $\{UCB(u)\}$ with temperature $\tau$.}
        $v \leftarrow u^\star$\;
        $\pi \leftarrow \pi \mathbin{\|} \langle v\rangle$\;
    }
    $s \leftarrow State(v)$ \tcp*{$s=(c,\mathcal{B})$}

    $(R,\mathit{diag},\textit{solved}) \leftarrow
    Evaluate(c,\,\mathcal{B}_{PM},\,\mathcal{B}_{CT};\,\alpha,\beta,\gamma)$\;
    \If{\textit{solved}}{
        \Return{$p_{\mathrm{sol}}\leftarrow c$}\;
    }
    $\mathcal{B} \leftarrow WriteBack(\mathcal{B},\mathit{diag})$\;
    $s \leftarrow (c,\mathcal{B})$\;
    \If{$R>R^{\mathrm{best}}$}{
        $R^{\mathrm{best}}\leftarrow R$\;
        $c^{\mathrm{best}}\leftarrow c$\;
    }

    $A \leftarrow \mathcal{A}_{\mathrm{str}}(s)\cup \mathcal{A}_{\mathrm{gen}}(s)\cup
                 \mathcal{A}_{\mathrm{test}}(s)\cup \mathcal{A}_{\mathrm{patch}}(s)$\;
    \ForEach{$a \in SelectSubset(A)$}{
        $s' \leftarrow T(s,a)$ \tcp*{$s'=(c',\mathcal{B}')$}
        AddChild$(v,NewNode(s'))$\;
    }

    \ForEach{$x \in \pi$}{
        $N(x) \leftarrow N(x) + 1$\;
        $\bar{X}(x) \leftarrow \dfrac{(N(x)-1)\bar{X}(x) + R}{N(x)}$\;
    }
}
\Return{$c^{\mathrm{best}}$}\;
\end{algorithm}

\subsection{Monte Carlo Tree Search as Global Planner}

To navigate the search space, we employ MCTS, a tree-based planning algorithm that incrementally constructs a search tree. Each node represents a state $s=(c,\mathcal{B})$, with edges corresponding to actions $a\in\mathcal{A}(s)$. MCTS explores the space through four canonical phases: selection, expansion, simulation, and backpropagation.

Given the search formulation in Section~\ref{sec:Problem Formulation as Search}, MCTS serves as a global planner that selects high-level actions $a \in \mathcal{A}(s)$ over states $s=(c,\mathcal{B})$, where $c$ is the current program draft and $\mathcal{B}$ is the blackboard. Each MCTS iteration follows the four canonical phases, while being adapted to competitive program generation: quality evaluation actions $a^{\text{eval}}\in\mathcal{A}_{\text{eval}}(s)$ are executed in the simulation/evaluation phase to produce both a scalar reward $R$ for backpropagation and structured diagnostics to update the blackboard $\mathcal{B}$. The detailed procedure is shown in Algorithm~\ref{alg:ariadne_mcts}.

\subsubsection{Selection}

Starting from the root state $s_0$, MCTS repeatedly selects a child node until reaching a node that is not fully expanded or is terminal. For each child $u$ of a node $v$, we compute the UCB score:

\begin{equation}
UCB(u) = \bar{X}(u) + C\sqrt{\frac{\ln(N(v)+1)}{N(u)}},
\label{eq:ucb}
\end{equation}
where $N(\cdot)$ and $\bar{X}(\cdot)$ denote visit counts and empirical mean values of the current node, and $C$ governs the trade-off between exploration and exploitation. We sample the next node from a softmax distribution over UCB scores with temperature $\tau$:

\begin{equation}
p(u)=\frac{\exp\big((UCB(u)-\max_w UCB(w))/\tau\big)}
{\sum_k \exp\big((UCB(k)-\max_w UCB(w))/\tau\big)}.
\label{eq:softmax}
\end{equation}
Here, $p(u)$ assigns higher probability to children with larger $UCB(u)$, while the temperature $\tau$ controls the sharpness of the selection distribution (i.e., smaller $\tau$ yields greedier choices, larger $\tau$ encourages more exploration). Subtracting $\max_w UCB(w)$ is a standard numerical-stability technique that does not alter the resulting distribution~~\cite{xu2024sra}. This stochastic selection mitigates over-commitment to a single branch and promotes diversified search trajectories, which is especially important when early feedback is noisy and multiple strategy hypotheses remain plausible.

\begin{figure*}[ht] 
    \centering 
    \includegraphics[width=1.0\textwidth]{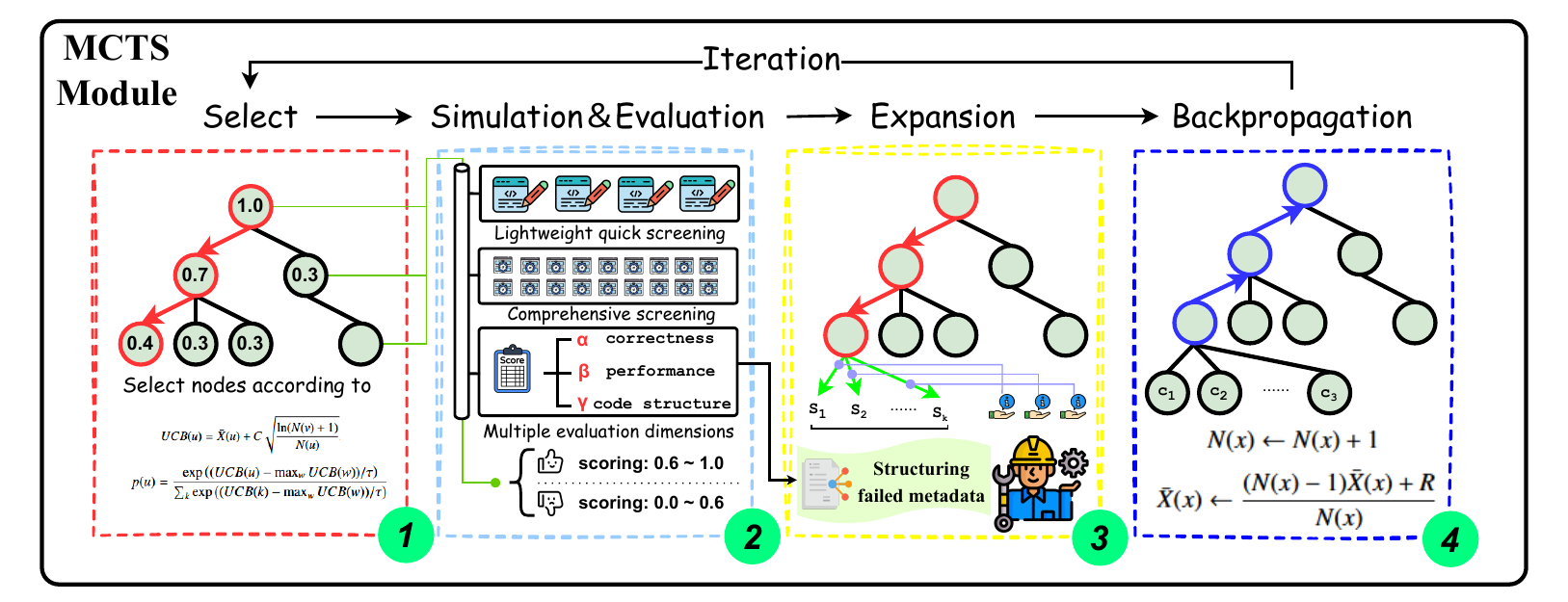}
\caption{Overview of the MCTS pipeline serving as a global planner, orchestrating agent actions through selection, expansion, simulation and evaluation, and backpropagation.} 
    \label{Fig:MCTS pipeline} 
\end{figure*}

\subsubsection{Simulation and Evaluation}

Given a selected node with state $s = (c, \mathcal{B})$, the current program draft $c$ is evaluated using a two-stage protocol to compute the node value and to extract diagnostic artifacts for updating the blackboard.

\noindent{\textbf{Two-stage evaluation.}}
We first perform a lightweight screening to filter out clearly invalid candidates and obtain coarse diagnostic information. If the draft passes this stage, a deeper evaluation is conducted using a more comprehensive test suite, with sandboxed execution applied when appropriate. A draft is considered a solution program $p_{\text{sol}}$ if it successfully passes the deep evaluation.

\noindent{\textbf{Reward computation.}}
For non-terminal drafts, we compute a scalar reward $R$ as a weighted combination of signals, following the value function $V(s)$ defined in Section~\ref{sec:Problem Formulation as Search}. We set $\alpha=0.6$ and $\beta=\gamma=0.2$ to make correctness the dominant driver of value estimates while still allowing efficiency and code-quality signals to break ties among drafts with similar test performance. In particular, when a draft achieves partial correctness but fails on specific counterexamples, the reward remains informative and encourages repair-focused actions that address the observed failures. Conversely, when correctness signals remain persistently low across attempts, the value estimates for the corresponding branches stay suppressed, prompting the search to explore alternative strategy branches rather than continuing local repairs. The evaluation additionally produces structured feedback, such as failing test cases and repair hints, which is recorded in $\mathcal{B}$ to guide subsequent action selection and state updates.

\noindent{\textbf{Reward computation.}}
For non-terminal drafts, we compute a scalar reward $R$ 
as a weighted combination of signals, following the value function $V(s)$ defined in Section~\ref{sec:Problem Formulation as Search}. The evaluation additionally produces structured feedback, such as failing test cases and repair hints, which is recorded in $\mathcal{B}$ to guide subsequent action selection and state updates.

\subsubsection{Expansion}

Unlike the canonical MCTS pipeline, which performs expansion before simulation, we conduct simulation and evaluation before expansion. Within the MCTS module, evaluation produces actionable artifacts that update the blackboard $\mathcal{B}$, thereby influencing which actions $a\in\mathcal{A}(s)$ are meaningful. Consequently, after evaluating a selected node, we enumerate high-level actions conditioned on the blackboard and apply a subset of them to generate successor states $s'=(c',\mathcal{B}')$, which are then added as child nodes.

\subsubsection{Backpropagation}

After obtaining the reward $R$ from evaluation, it is backpropagated along the selected path to update visit counts and value estimates:
\begin{equation}
N(x) \leftarrow N(x) + 1,\qquad
\bar{X}(x) \leftarrow \frac{(N(x)-1)\bar{X}(x)+R}{N(x)},
\end{equation}
for every node $x$ along the trajectory. These statistics are subsequently utilized by the UCB rule in Eq.~\ref{eq:ucb} to guide future action selection.

\subsection{Agent Design}
\label{sec:agentdesign}

Competitive program generation is inherently iterative: a solver should (i) propose multiple candidate algorithmic strategies, (ii) implement the selected strategy as an executable program, (iii) validate correctness by running targeted test cases, and (iv) repair failures based on concrete evidence. Under our search formulation, each MCTS node corresponds to a state $s=(c,\mathcal{B})$, consisting of a program draft $c$ and a shared blackboard $\mathcal{B}$, where each tree edge corresponds to an action $a\in\mathcal{A}(s)$ that transforms the state according to $s' = T(s,a)$. The action space $\mathcal{A}(s)$ is implemented via a set of specialized agents.

\begin{figure*}[ht] 
    \centering 
     \includegraphics[width=0.8\textwidth]{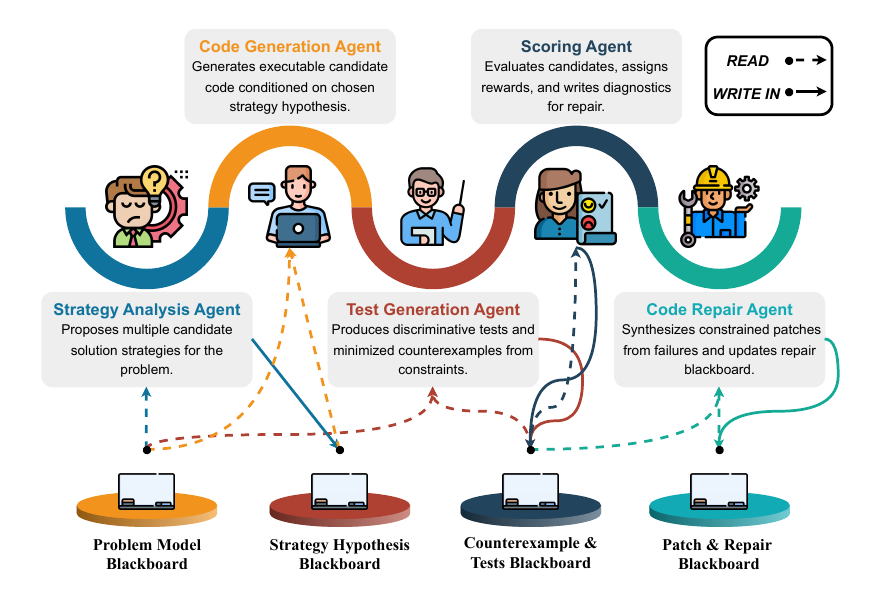}
\caption{Overview of agent–blackboard information exchange, where agents read structured entries from the blackboard and write back reusable artifacts that guide subsequent search actions.} 
    \label{Fig:agent design} 
\end{figure*}

Each agent instantiates a specific space of actions within $\mathcal{A}(s)$  according to a fixed interaction contract: it reads structured entries from designated blackboard components, performs a bounded reasoning or synthesis procedure, and writes back reusable artifacts. This design renders the search evidence-driven: quality evaluation actions not only produce a scalar reward $R$ for backpropagation but also generate structured diagnostics that update $\mathcal{B}$, thereby conditioning subsequent generation and repair actions. Figure~\ref{Fig:agent design} illustrates the resulting closed-loop information flow, and the full prompt templates for all agents are provided in ~\ref{app:prompts}.

\subsubsection{Strategy Analysis Agent}

The Strategy Analysis Agent \texorpdfstring{$\mathcal{A}_{\text{str}}$}{(A-str)} instantiates \textbf{strategy selection actions} $a^{\text{str}}\in \mathcal{A}_{\text{str}}(s)$. Given a state $s=(c,\mathcal{B})$, it reads the structured task specification from the Problem Model Blackboard and proposes a set of algorithmic strategy hypotheses. Each hypothesis is associated with an estimated utility, such as expected correctness likelihood and computational cost, forming a strategy prior that is recorded on the Strategy Hypothesis Blackboard. This prior subsequently conditions actions in $\mathcal{A}_{\text{gen}}(s)$ and $\mathcal{A}_{\text{patch}}(s)$, biasing the search toward promising algorithmic trajectories. The full prompt templates are shown in ~\ref{app:prompts_strategy}.

\subsubsection{Code Generation Agent}
The Code Generation Agent \texorpdfstring{$\mathcal{A}_{\text{gen}}$}{(A-gen)} instantiates \textbf{code generation actions} $a^{\text{gen}}\in \mathcal{A}_{\text{gen}}(s)$. It reads the active strategy hypothesis and the problem model from $\mathcal{B}$ to produce an updated draft program $c'$. Formally, it realizes the transition $T\big((c,\mathcal{B}),a^{\text{gen}}\big)$ = $(c',\mathcal{B}')$, where $\mathcal{B}'$ records the strategy commitment metadata for attribution and subsequent hypothesis refinement. The full prompt templates are shown in ~\ref{app:prompts_codegen}.

\subsubsection{Test Generation Agent}
The Test Generation Agent \texorpdfstring{$\mathcal{A}_{\text{test}}$}{(A-test)} instantiates \textbf{test construction actions} $a^{\text{test}}\in \mathcal{A}_{\text{test}}(s)$. 
It reads constraints and invariants from the Problem Model Blackboard, as well as accumulated evidence (tests and counterexamples) from the Counterexample \& Tests Blackboard, and generates a prioritized test set within a constrained execution budget. The resulting transition updates $\mathcal{B}$ by adding discriminative tests and minimizing counterexamples, thereby improving screening efficiency and failure localization for subsequent evaluation and repair. The full prompt templates are shown in ~\ref{app:prompts_testgen}.

\subsubsection{Scoring Agent}
The Scoring Agent \texorpdfstring{$\mathcal{A}_{\text{eval}}$}{(A-eval)} instantiates \textbf{quality evaluation actions} $a^{\text{eval}}\in \mathcal{A}_{\text{eval}}(s)$ and serves as the concrete implementation of the value and reward definitions. Given a candidate draft $c$ and the blackboard context $\mathcal{B}$, it performs staged testing using the curated test pool and produces (i) a scalar reward $R$ for MCTS backpropagation and (ii) structured diagnostics such as failing inputs, violated conditions, and error traces. The diagnostics are written back to the Counterexample \& Tests Blackboard and the Patch \& Repair Blackboard, enabling the reuse of evaluation evidence across search branches. The full prompt templates are shown in ~\ref{apps:prompts_scoring}.

\subsubsection{Code Repair Agent}
The Code Repair Agent \texorpdfstring{$\mathcal{A}_{\text{patch}}$}{(A-patch)} instantiates \textbf{patch application actions} $a^{\text{patch}}\in \mathcal{A}_{\text{patch}}(s)$. It reads failure evidence from the Counterexample \& Tests Blackboard and repair-related signals from the Patch \& Repair Blackboard, and then proposes constrained patch objects characterized by applicability conditions and compatibility constraints. During expansion, the coordinator selects compatible patches and applies them to produce repaired drafts $c'$, resulting in transitions of the form $T\big((c,\mathcal{B}),a^{\text{patch}}\big)$ = $(c',\mathcal{B}')$, where $\mathcal{B}'$ records the applied patch and the updated repair state. The full prompt templates are shown in ~\ref{app:prompts_repair}.

\subsection{Blackboard System}
\label{sec:blackboard}

Competitive program generation differs from standard one-shot code synthesis in two fundamental aspects: (i) correctness is primarily verified through test-driven evidence rather than purely textual reasoning, and (ii) the solution process is inherently iterative, alternating among strategy proposal, code generation, testing, and repair. Under our search formulation, each node state is represented as $s=(c,\mathcal{B})$, where $c$ denotes the current program draft and $\mathcal{B}$ represents a shared blackboard. The introduction of the blackboard serves two purposes. First, it provides a persistent and structured workspace that accumulates constraints, hypotheses, counterexamples, and repair cues discovered during search, preventing intermediate artifacts from being discarded between rollouts. Second, it facilitates cross-branch reuse: evidence generated during the evaluation of one draft (e.g., a failing input) can directly inform subsequent actions applied to other drafts, thereby reducing redundant exploration and accelerating convergence toward a correct solution.

\begin{figure}[!htbp]
    \centering 
    \includegraphics[width=0.45\textwidth]{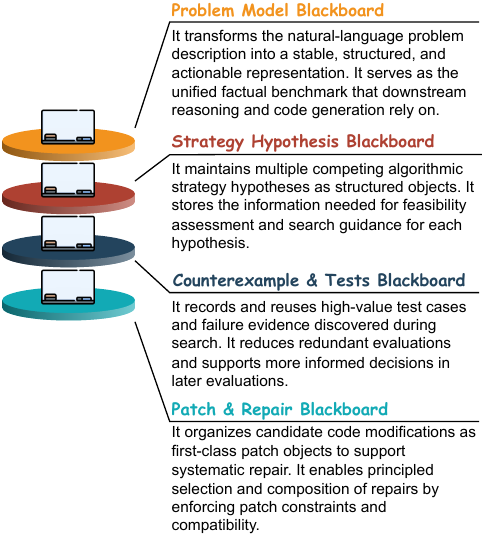}
\caption{Structural organization of the blackboard system in {\tool}.} 
    \label{Fig:blackboard}
\end{figure}

We further decompose $\mathcal{B}$ into multiple dedicated components, as the intermediate artifacts in competitive programs are heterogeneous and exhibit different lifecycles and access patterns. For instance, the normalized problem semantics should remain stable once constructed, whereas failure evidence and repair suggestions evolve continuously as the search explores new drafts. Storing these artifacts in an unstructured memory would lead to inconsistent interpretations, redundant information, and ineffective conditioning of actions. To address this, we organize the blackboard into four components, $\mathcal{B}=(\mathcal{B}_{PM},\mathcal{B}_{SH},\mathcal{B}_{CT},\mathcal{B}_{PR})$, each aligned with a distinct stage of the generate–test–repair loop and directly consumed by the corresponding action families in $\mathcal{A}(s)$. Figure~\ref{Fig:blackboard} illustrates this organization and the information flow among components.

\subsubsection{Problem Model Blackboard}
\label{sec:problem_model_blackboard}

 \texorpdfstring{($\mathcal{B}_{PM}$)}{(B-PM)}

The Problem Model Blackboard $\mathcal{B}_{PM}$ provides a stable and structured representation of the problem instance, serving as the shared semantic reference for all actions. Its primary function is to translate the natural-language problem statement into canonical fields that can be directly consumed by strategy selection ($\mathcal{A}_{\text{str}}$), code generation ($\mathcal{A}_{\text{gen}}$), test generation ($\mathcal{A}_{\text{test}}$), evaluation ($\mathcal{A}_{\text{eval}}$), and repair ($\mathcal{A}_{\text{patch}}$).

\noindent\textbf{Canonical fields.}
$\mathcal{B}_{PM}$ stores four categories of structured entries:
\begin{itemize}
    \item \textbf{Constraints:} explicit specifications of input ranges and resource limits (e.g., time and memory).
    \item \textbf{Objective specification:} a normalized description of the required output behavior and optimization objectives, if applicable.
    \item \textbf{Required invariants:} semantic properties that any correct solution must satisfy and preserve.
    \item \textbf{Edge-case checklist:} a systematically enumerated set of boundary and degenerate cases used to ensure robustness.
\end{itemize}

\noindent\textbf{Normalization.}
To ensure consistency across search branches, the extracted entries are canonicalized into a non-redundant form by normalizing numeric formats, range expressions, variable names, and equivalent constraints. This canonicalization prevents duplicated or inconsistent interpretations of the problem and enables downstream actions to operate on a unified semantic reference.

\noindent\textbf{Usage in search.}
By centralizing constraints, invariants, and edge cases within $\mathcal{B}_{PM}$, the search process reduces unnecessary exploration. Invalid strategy hypotheses can be deprioritized early, generated code can be verified against explicit requirements, and test generation can systematically target boundary conditions. This design enhances both the efficiency of the search and the reliability of the generated programs.

\subsubsection{Strategy Hypothesis Blackboard}
\label{sec:strategy-hypothesis-blackboard}

 \texorpdfstring{($\mathcal{B}_{SH}$)}{(B-SH)}

The Strategy Hypothesis Blackboard $\mathcal{B}_{SH}$ maintains multiple competing strategy hypotheses as structured artifacts, enabling strategy selection to be explored dynamically during the search rather than fixed a priori. It is primarily produced by strategy selection actions ($\mathcal{A}_{\text{str}}$) and consumed by code generation actions $\mathcal{A}_{\text{gen}}$ and repair actions $\mathcal{A}_{\text{patch}}$.
In addition, $\mathcal{B}_{SH}$ can materialize a lightweight requirements document derived from $\mathcal{B}_{PM}$, which summarizes problem constraints, target invariants, edge cases, and performance requirements. This document is provided to the Strategy Analysis Agent, which generates multiple diverse candidate strategies (e.g., alternative algorithmic paradigms or proof sketches) rather than a single committed plan. The resulting strategies are written back to $\mathcal{B}_{SH}$ as competing hypotheses to be explored and refined throughout the search process.

\noindent\textbf{Stored hypothesis structure.}
Each strategy hypothesis is stored along with:
(1) applicability conditions derived from $\mathcal{B}_{PM}$;
(2) an estimated upper bound on time and memory complexity;
(3) risk annotations indicating likely failure modes (e.g., TLE, overflow, off-by-one errors); and
(4) a concise rationale linked to the required invariants from $\mathcal{B}_{PM}$.

\noindent\textbf{Prioritization for search.}
When multiple hypotheses coexist, $\mathcal{B}_{SH}$ stores normalized priority weights (priors) to bias action selection toward more promising strategy directions. Specifically, $\mathcal{B}_{SH}$ ranks strategies according to a predefined set of criteria (e.g., feasibility under constraints from $\mathcal{B}_{PM}$, asymptotic complexity bounds, implementation difficulty, and estimated risk), and converts this ranking into priors that guide subsequent action choices. These weights are dynamically updated as quality evaluation actions ($\mathcal{A}_{\text{eval}}$) generate new evidence supporting or contradicting the current hypothesis set. Through this mechanism, nodes derived from higher-ranked strategies are more likely to achieve superior empirical returns during evaluation and, consequently, receive higher node values via backpropagation, further increasing their selection probability in later search iterations.

\subsubsection{Counterexample \& Tests Blackboard}
\label{sec:counterexample_blackboard}

 \texorpdfstring{($\mathcal{B}_{CT}$)}{(B-CT)}

The Counterexample \& Tests Blackboard $\mathcal{B}_{CT}$ records reusable test artifacts and failure evidence discovered during the search. In competitive programming, tests provide the most direct and discriminative feedback for distinguishing correct implementations from plausible but incorrect ones. By persisting high-impact tests and counterexamples in $\mathcal{B}_{CT}$, the system avoids repeatedly rediscovering the same failure modes across different search branches.

\noindent\textbf{Stored artifacts.}
$\mathcal{B}_{CT}$ stores:
(1) generated stress tests and boundary inputs derived from $\mathcal{B}_{PM}$;
(2) failing inputs (counterexamples) and their minimized forms; and
(3) structured failure descriptors, such as WA, RE, or TLE patterns and associated execution traces when available.

\noindent\textbf{Integration with evaluation.}
Quality evaluation actions ($\mathcal{A}_{\text{eval}}$) consult $\mathcal{B}_{CT}$ to perform low-cost screening and to identify discriminative counterexamples. Newly discovered failures are written back to $\mathcal{B}_{CT}$, thereby making subsequent evaluations both more efficient and more informative.

\subsubsection{Patch \& Repair Blackboard}
\label{sec:patch_repair_blackboard}

 \texorpdfstring{($\mathcal{B}_{PR}$)}{(B-PR)}

The Patch \& Repair Blackboard $\mathcal{B}_{PR}$ organizes repair suggestions as structured artifacts that can be reused and composed throughout the search. It is primarily produced by repair actions ($\mathcal{A}_{\text{patch}}$) based on failure evidence from $\mathcal{B}_{CT}$ and constraints/invariants from $\mathcal{B}_{PM}$, and it constrains the repair-related portion of the action space during expansion. In competitive program generation, many failures are not resolved by regenerating code from scratch; instead, they require targeted modifications guided by concrete counterexamples. By storing repairs as explicit objects, $\mathcal{B}_{PR}$ enables the search to systematically explore \emph{repair trajectories} rather than performing ad hoc mutations.

\noindent\textbf{Patch representation.}
Each entry on $\mathcal{B}_{PR}$ corresponds to a candidate patch represented independently of any specific code instance, and is annotated with: 
(1) the targeted failure pattern linked to $\mathcal{B}_{CT}$; 
(2) applicability conditions derived from $\mathcal{B}_{PM}$; and 
(3) the expected effect on the program draft. 

To reflect different modification scopes and associated costs, patches are organized into three granularity levels:
\begin{itemize}
    \item \textbf{L1 (Local Repairs):} Small, low-cost edits that correct localized implementation issues (e.g., boundary handling or indexing errors) without altering the overall program structure.
    \item \textbf{L2 (Structural Repairs):} Medium-scale refactorings that preserve the algorithmic strategy but modify internal organization (e.g., changing data structures or loop constructs).
    \item \textbf{L3 (Strategy Repairs):} High-level modifications that revise or replace the underlying strategy hypothesis (e.g., switching to an alternative hypothesis stored in $\mathcal{B}_{SH}$).
\end{itemize}

\noindent\textbf{Compatibility constraints.}
To prevent incoherent repair compositions, $\mathcal{B}_{PR}$ maintains metadata indicating when patches are incompatible or when one patch should precede another. Specifically, each patch may include: 
(1) a set of preconditions (e.g., an overflow-related diagnostic must be present); 
(2) a conflict set of mutually exclusive patches (e.g., incompatible indexing conventions); and 
(3) ordering constraints for prerequisite transformations (e.g., preprocessing must be applied before a dependent data-structure modification). 
During expansion, repair actions are filtered according to these constraints, ensuring that MCTS explores only valid repair combinations.

\noindent\textbf{Role in search transitions.}
Applying a patch corresponds to executing an action $a^{\text{patch}} \in \mathcal{A}_{\text{patch}}(s)$, which produces a transition 
$T((c, \mathcal{B}), a^{\text{patch}}) = (c', \mathcal{B}')$.
The applied patch and its associated metadata are recorded back into $\mathcal{B}_{PR}$, while newly observed outcomes (e.g., whether a counterexample is resolved) are written into $\mathcal{B}_{CT}$. By storing repair artifacts in $\mathcal{B}_{PR}$ and conditioning expansion on their applicability, the system enables evidence-driven refinement of code drafts while keeping the repair search space tractable.

\section{Experimental Setup}
\label{sec:setup}

\subsection{Research Questions}

To systematically evaluate the effectiveness of our proposed approach, we design the following four research questions (RQs) to guide our empirical evaluation.

\textbf{RQ1: Does our proposed approach outperform baseline methods on competitive program generation?}

\textbf{Design Motivation.} Previous multi-agent code generation frameworks~\cite{islam2024mapcoder,islam2025codesim,chen2025treemind,pan2025codecor,li2025codetree,zhang2024codeagent,huang2023agentcoder} exhibit two primary limitations: (1) their workflows are often fixed, with agents invoked in a predetermined linear sequence, lacking dynamic decision-making and global coordination informed by intermediate results; and (2) they focus primarily on generating syntactically plausible code or satisfying simple checks, rather than systematically exploring and optimizing within a structured solution space.
To conduct a comprehensive assessment, we compared our method against four state-of-the-art baselines. 
Specifically, MapCoder~\cite{islam2024mapcoder} simulates step-by-step human reasoning to coordinate agent collaboration for final code generation. 
CodeSim~\cite{islam2025codesim} employs a three-stage pipeline with an adaptive loop of planning, generation, and refinement. 
CodeTree~\cite{li2025codetree} organizes generation and evaluation in a tree structure, expanding or pruning branches based on execution feedback until convergence.

\textbf{RQ2: Does the MCTS-based strategy in our framework enable more effective exploration than alternative search strategies?}

\textbf{Design Motivation.} 
For search tasks within a large-scale solution space, alternative strategies such as Breadth-First Search (BFS), Depth-First Search (DFS), and greedy heuristics can also traverse candidate solutions according to their respective search logic. However, these strategies often struggle to balance exploration and exploitation effectively: they may either explore too narrowly, missing promising regions of the solution space, or explore too broadly, incurring high computational cost without sufficient guidance. 
In this RQ, we aim to perform a comparative ablation study to evaluate whether the MCTS-based strategy in our framework enables more effective exploration than these alternatives, in terms of solution quality, search efficiency, and robustness, while operating under the same blackboard-driven agent architecture.

\textbf{RQ3: How do the key hyperparameters in MCTS influence program generation performance and search efficiency?}

\textbf{Design Motivation.} 
The performance of MCTS is highly sensitive to its key hyperparameters. Limiting the maximum search depth helps control computational resources, but a depth that is too shallow may fail to explore complex repair sequences, whereas an excessively deep tree can lead to a combinatorial explosion and waste resources on unproductive paths. Similarly, the exploration–exploitation balance, governed by the UCB coefficient ($C$) and the temperature hyperparameter ($\tau$) used in probabilistic action selection, critically affects search behavior: improper tuning can cause premature convergence to local optima or inefficient wandering in low-quality regions.
Therefore, we aim to quantitatively evaluate the sensitivity of these hyperparameters and identify configurations that optimize both performance and efficiency for competitive program generation, providing empirical guidance for tuning {\tool} in practical deployments.

\textbf{RQ4: How effective is our proposed approach under realistic competitive programming contests?}

\textbf{Design Motivation.} While controlled benchmarks provide a useful means for evaluating algorithmic correctness and relative performance, they do not fully capture the challenges encountered in real competitive programming contests. In practice, solutions must be developed under strict time constraints, limited computational resources, and incomplete feedback, while generalizing across a diverse set of problem types. Human competitors must continuously balance solution quality, implementation efficiency, and time management, making competitive programming a fundamentally dynamic and resource-constrained setting.
To evaluate whether our proposed approach can operate effectively in such realistic contests, we assess its performance under constrained time budgets, varying problem difficulties, and limited interaction opportunities, comparing it against both human-level and automated baselines. Beyond absolute success rates, we consider multiple performance dimensions, including solution correctness, computational cost, and resource utilization.

\subsection{Experimental Subjects}

To address the first three research questions, we evaluate our approach on four representative competitive programming benchmarks, encompassing diverse problem domains, difficulty levels, and evaluation protocols.

\begin{itemize}
    \item \textbf{CodeContests}~\cite{li2022competition} is a large-scale benchmark comprising competitive programming problems collected from multiple online judges, each paired with executable test cases for automatic evaluation. It provides a standardized setting for assessing algorithmic reasoning across a wide range of classical programming tasks.
    \item \textbf{CodeContests+}~\cite{wang2025codecontests+} extends CodeContests by replacing the original test suites with higher-quality, automatically generated and validated tests. This enhanced version improves the reliability of correctness evaluation and mitigates false positives arising from insufficient or weak test coverage.
    \item \textbf{APPS}~\cite{hendrycks2021measuring} is a large-scale program synthesis benchmark encompassing problems of varying difficulty, from introductory exercises to complex algorithmic challenges. Solutions are evaluated through executable test cases, making it a widely adopted benchmark for general-purpose code generation capability.
    \item \textbf{LiveCodeBench}~\cite{jain2024livecodebench}
    aims to provide a more comprehensive and contamination-free assessment of coding capabilities by continuously collecting newly released problems from periodic contests on major competitive-programming platforms.
\end{itemize}

To address \textbf{RQ4}, we evaluate our approach under realistic competitive programming conditions using complete contest instances collected from Codeforces. Each instance corresponds to a full ICPC-style contest, comprising multiple problems released simultaneously under a fixed time limit.

\subsection{Performance Metrics}

Following prior work~\cite{zan2022cert,zheng2023codegeex}, we adopt pass rate as the primary evaluation metric. 
A generated program is considered correct only if it successfully passes all test cases in the evaluation suite. 
In particular, we focus on the Pass@1 metric~\cite{islam2024mapcoder}, which evaluates whether a single generated solution is correct. This choice reflects practical usage scenarios in which only one final program is typically produced and submitted for evaluation.

\subsection{Baselines}

We employ three representative baseline methods to evaluate the performance of our proposed approach:

\begin{itemize}

    \item \textbf{MapCoder.} MapCoder~\cite{islam2024mapcoder} is a multi-agent framework designed to emulate the step-by-step reasoning process typically employed by human programmers. It decomposes code generation into a sequence of coordinated reasoning stages, in which specialized agents collaboratively analyze the problem, plan solution strategies, and generate code. By explicitly modeling the human problem-solving workflow, MapCoder seeks to improve solution quality through structured reasoning rather than relying on one-shot generation.

    \item \textbf{CodeSim.} CodeSim~\cite{islam2025codesim} is a multi-agent framework that employs a staged pipeline comprising planning, generation, and refinement phases. Agents interact through an iterative loop in which high-level planning guides code generation, and subsequent refinement steps correct errors or enhance solution quality. This coordination mechanism allows CodeSim to progressively refine candidate programs based on intermediate feedback.

    \item \textbf{CodeTree.} CodeTree~\cite{li2025codetree} formulates program generation as a tree-structured search problem. Starting from an initial root node, it incrementally expands candidate solution nodes and evaluates them using execution-based feedback. Based on these evaluations, the framework dynamically prunes unpromising branches while further exploring promising ones, gradually converging toward high-quality solutions.
\end{itemize}

\subsection{Selection of LLMs}

We evaluate four representative LLMs, including DeepSeek-v3.2, GPT-4o, Qwen3-Coder-480B, and Gemini-2.5-flash-thinking, which collectively represent general-purpose, code-specialized, and reasoning-oriented models, thus reflecting the dominant deployment scenarios of current large language model systems.
Specifically, \textbf{DeepSeek-v3.2} and \textbf{GPT-4o} serve as representative general-purpose LLMs, providing strong natural language understanding, instruction following, and broad reasoning capabilities across diverse tasks. \textbf{Qwen3-Coder-480B} represents coder-oriented LLMs, emphasizing program synthesis, implementation correctness, and engineering-focused behaviors, which are particularly relevant for competitive program generation. \textbf{Gemini-2.5-flash-thinking} is included as a reasoning-oriented model that explicitly prioritizes multi-step deliberation, allowing us to evaluate our method under a reasoning-centric generation regime.

\subsection{Implementation Details and Running Platform}
Our framework is implemented by invoking multiple mainstream LLM APIs via the unified gateway provided by \texttt{Yunwu API}\footnote{\url{https://yunwu.ai/}}, which ensures consistent request formatting, authentication, and logging across different providers, and facilitates seamless switching between model endpoints during experimentation.

All experiments were conducted on a high-performance workstation equipped with an Intel Core i7-13600K CPU, 32 GB of RAM, and an NVIDIA GeForce RTX 4090 GPU with 24 GB of memory, running Windows 10.

\section{Result Analysis}
\label{sec:results}

\subsection{RQ1: Comparison with Baselines}

\textbf{Approach.} 
To evaluate the performance of our proposed competitive program generation framework, {\tool}, we compare it against three state-of-the-art baselines, including MapCoder~\cite{islam2024mapcoder}, CodeSim~\cite{islam2025codesim}, and CodeTree~\cite{li2025codetree}. These baselines employ diverse multi-agent collaboration strategies, providing a comprehensive assessment of {\tool}'s effectiveness. We primarily adopt the Pass@1 metric~\cite{zan2022cert,zheng2023codegeex} to measure solution correctness. Additionally, we record runtime statistics (such as execution time, token consumption, and Monte Carlo search tree depth) to offer a more complete evaluation of performance and efficiency.

\begin{table*}[!htbp]
\centering
\caption{Performance comparison of {\tool} and state-of-the-art baselines, reported in Pass@1 (\%).}
\label{tab:comparison}
\begin{tabular}{ll
  >{\centering\arraybackslash}p{2cm}
  >{\centering\arraybackslash}p{2cm}
  >{\centering\arraybackslash}p{2cm}
  >{\centering\arraybackslash}p{2cm}}
\toprule
\multirow{2}{*}{LLM} & \multirow{2}{*}{Method} & \multicolumn{4}{c}{Dataset (pass@1 \%)} \\
\cmidrule(lr){3-6}
 &  & APPS & CodeContest & CodeContest+ & LiveCodeBench \\
\midrule

\multirow{4}{*}{DeepSeek-V3.2}
 &  {\tool} & 44.67 & 42.42 & 23.64 & 22.73 \\
 & MapCoder  & 14.00 & 26.06 & 10.00 & 8.18 \\
 & CodeSim   & 17.33 & 24.85 & 12.72 & 15.45 \\
 & CodeTree  & 16.00 & 25.45 & 11.82 & 12.73 \\
\midrule

\multirow{4}{*}{GPT-4o}
 &  {\tool} & 41.30 & 46.67 & 27.27 & 20.91 \\
 & MapCoder  & 12.67 & 20.61 & 8.18 & 12.73 \\
 & CodeSim   & 18.00 & 26.06 & 14.55 & 11.82 \\
 & CodeTree  & 15.33 & 23.03 & 10.91 & 11.82 \\
\midrule

\multirow{4}{*}{Qwen3-Coder-480B}
 &  {\tool} & 27.33 & 32.12 & 22.73 & 14.55 \\
 & MapCoder  & 14.67 & 18.79 & 6.36 & 9.09 \\
 & CodeSim   & 19.33 & 23.64 & 13.64 & 12.73 \\
 & CodeTree  & 16.67 & 21.21 & 10.00 & 10.91 \\
\midrule

\multirow{4}{*}{Gemini-2.5-Flash-Thinking}
 &  {\tool} & 27.33 & 29.09 & 26.36 & 26.36 \\
 & MapCoder  & 12.67 & 12.72 & 10.91 & 12.73 \\
 & CodeSim   & 18.00 & 20.00 & 16.36 & 18.18 \\
 & CodeTree  & 14.67 & 18.18 & 13.64 & 14.55 \\
\bottomrule
\end{tabular}
\end{table*}

\textbf{Result.} 
Table~\ref{tab:comparison} compares {\tool} against three strong baselines (MapCoder, CodeSim, and CodeTree) under matched settings across four LLM backbones and four benchmarks (16 configurations in total). In every configuration, {\tool} achieves the highest Pass@1, indicating that its gains are not tied to a specific model or dataset. Aggregated over all configurations, {\tool} attains an average Pass@1 of 29.72\%, substantially higher than the best baseline CodeSim (17.67\%), corresponding to a 68.2\% relative improvement. The advantage is particularly clear on APPS with GPT-4o, where {\tool} reaches 41.3\% (62/150 solved), yielding relative improvements of 225.97\% over MapCoder, 129.44\% over CodeSim, and 169.41\% over CodeTree, and similar dominance holds on CodeContests, CodeContests+, and LiveCodeBench.

From the perspective of LLM backbones, {\tool} remains consistently effective while absolute performance varies with the underlying model. Averaged over the four datasets, {\tool} achieves 33.37\% (DeepSeek-V3.2), 34.04\% (GPT-4o), 24.18\% (Qwen3-Coder-480B), and 27.29\% (Gemini-2.5-Flash-Thinking), outperforming the strongest baseline under each backbone by 89.7\%, 93.3\%, 39.5\%, and 50.5\%, respectively. This pattern suggests that {\tool} provides robust search-and-repair benefits that transfer across model families rather than exploiting idiosyncrasies of a single LLM.

Finally, comparing across datasets reveals that {\tool}'s improvements persist under different domains and evaluation protocols. Relative to the strongest baseline averaged across backbones, {\tool} improves Pass@1 by 93.5\% on APPS, 59.0\% on CodeContests, 74.6\% on CodeContests+, and 45.3\% on LiveCodeBench. Notably, although CodeContests+ and LiveCodeBench generally yield lower absolute Pass@1 for all methods, {\tool} maintains a clear lead, indicating better generalization and higher exploration efficiency in more challenging or less familiar problem distributions.

The observed superiority of {\tool} can be attributed to several factors. First, the blackboard system promotes precise problem formulation, reducing misinterpretation by agents. Second, it maintains contextual consistency across multi-agent interactions while persistently updating key evidence to guide code refinement. Third, unlike baselines with fixed collaboration workflows, {\tool} dynamically invokes appropriate agents according to the evolving code state, enhancing flexibility. Finally, the MCTS transforms code generation into an iterative process, effectively balancing exploration of alternative solutions with exploitation of high-value drafts.

\begin{tcolorbox}[width=1.0\linewidth, title={Answer to RQ1:}]

{\tool} consistently outperforms baselines in Pass@1 across all four benchmarks and multiple LLM backends. The highest overall performance is observed with DeepSeek-V3.2, achieving 44.67 on APPS and 42.42 on CodeContests. These results demonstrate that the integration of a blackboard-driven MCTS framework substantially enhances competitive program generation, improving first-attempt correctness and overall solution reliability.

\end{tcolorbox}

\subsection{RQ2: Comparison of different search strategies}

\textbf{Approach.} 
In RQ2, we investigate the impact of different search strategies on {\tool}’s ability to identify optimal program solutions. 
We compare the MCTS planner with three standard baselines: Breadth-First Search (BFS), Depth-First Search (DFS), and Greedy Heuristic Search\footnote{Greedy heuristic search selects the locally best action at each step without backtracking; for comparability, it is modeled as a bounded single-path search analogous to a degenerate tree.}. All strategies are evaluated on identical problem instances, employ the same LLM backends and prompt templates, and share the same agent interfaces and execution environment. Each algorithm follows the same evaluation protocol and reward computation, differing solely in the search policy guiding action selection and expansion. We report \textit{pass@1} as the primary metric, alongside runtime and token consumption. To assess efficiency under resource constraints, we further report token-normalized \textit{pass@1}, defined as \textit{pass@1} divided by total token usage.

\textbf{Result.} 
Table~\ref{tab:search_strategies} compares \textit{pass@1} across different search strategies (MCTS, BFS, DFS, Greedy) under the same blackboard-driven agent architecture and identical inference budgets on four datasets (APPS, CodeContests, CodeContests+, LiveCodeBench). MCTS consistently outperforms all baselines, achieving \textit{pass@1} scores of 41.33 (APPS), 46.67 (CodeContests), 27.27 (CodeContests+), and 20.91 (LiveCodeBench). BFS and DFS show moderate performance (e.g., 37.33 and 36.67 on representative datasets), whereas Greedy performs worst (e.g., 14.55 on LiveCodeBench), highlighting the limitations of na\"ive or purely local search strategies.

The superior performance of MCTS can be attributed to its principled balance between exploration and exploitation. Unlike Greedy, which risks early commitment to locally optimal paths, MCTS maintains diverse exploration across plausible solution trajectories. Compared with DFS, which may over-invest in a single branch, MCTS iteratively refines node-value estimates via backpropagation, enabling globally informed decisions. Compared with BFS, which spreads computation uniformly and may dilute the search budget over low-value nodes, MCTS reallocates expansions toward empirically promising nodes, improving sample efficiency. When combined with the blackboard’s persistent intermediate artifacts and feedback signals, MCTS more effectively consolidates information to guide search, resulting in consistently higher \textit{pass@1} across all datasets.

\begin{table*}[!htbp]
\centering
\caption{Comparison of search strategies (MCTS, BFS, DFS, and Greedy) under the same blackboard-driven agent architecture, reported in Pass@1 (\%).}
\label{tab:search_strategies}
\begin{tabular}{l >{\centering\arraybackslash}p{3cm} >{\centering\arraybackslash}p{3cm} >{\centering\arraybackslash}p{3cm} >{\centering\arraybackslash}p{3cm}}
\toprule
\textbf{Search Strategy} & \textbf{APPS} & \textbf{CodeContests} & \textbf{CodeContests+} & \textbf{LiveCodeBench} \\
\midrule

\textbf{ {\tool}-MCTS}   & 41.33 & 46.67 & 27.27 & 20.91 \\

 {\tool}-BFS    & 37.33 & 41.21 & 23.64 & 17.27 \\

 {\tool}-DFS    & 36.67 & 40.00 & 22.73 & 16.36 \\

 {\tool}-Greedy & 34.67 & 36.36 & 20.00 & 14.55 \\
\bottomrule
\end{tabular}
\end{table*}

\begin{tcolorbox}[width=1.0\linewidth, title={Answer to RQ2:}]

Using MCTS as the planner enhances {\tool}'s search effectiveness within the same blackboard-driven agent architecture by adaptively balancing exploration and exploitation. Unlike BFS or DFS, which either diffuse the search budget across many branches or over-commit to a single path, MCTS concentrates computation on high-potential action sequences while preserving diversity, mitigating premature convergence.

\end{tcolorbox}

\subsection{RQ3: Influence of Key Hyperparameters}

\textbf{Approach.}
We jointly tune three key MCTS hyperparameters: the maximum tree depth $D$, the UCB exploration coefficient $C$ (Eq.~\ref{eq:ucb}), and the softmax temperature $\tau$ (Eq.~\ref{eq:softmax}) used for child-node sampling. To isolate the effect of the search policy, we fix the problem set, model backend, inference settings, prompt templates, tooling/sandbox, and the quick/deep evaluation protocol with a consistent budget across all runs. All experiments in this section are conducted on the CodeContests dataset (165 problems). As a reference, we compare against a \emph{default MCTS baseline} $\theta_{\mathrm{base}}=(D_{\mathrm{base}}=4,\, C_{\mathrm{base}}=1,\, \tau_{\mathrm{base}}=0.6)$, which serves as the standard configuration throughout this study. The selected model for this analysis is \textbf{DeepSeek-V3.2}, identified in RQ1 as the top-performing LLM.

Since MCTS hyperparameters affect both effectiveness and efficiency, tuning $(D, C,\tau)$ naturally induces trade-offs between solution quality and computational cost. We therefore cast hyperparameter selection as a multi-objective optimization problem: for each configuration $\theta=(D, C,\tau)$, we seek to maximize success rate (Pass@1) while minimizing resource consumption in terms of tokens and wall-clock time. Rather than collapsing these metrics into a single scalar with arbitrary weights, we first identify configurations that are Pareto-efficient with respect to accuracy and efficiency, and then apply a deterministic tie-breaking rule that prioritizes Pass@1, followed by tokens-per-solved and time-per-solved.

Due to computational constraints, we adopt a two-stage joint hyperparameter selection procedure. In Stage~1 (coarse screening), we evaluate a sparse factorial grid over $(D, C, \tau)$, running each configuration with a limited number of random seeds. For each configuration $\theta=(D, C,\tau)$, we record \textit{pass@1} $p(\theta)$, total token usage $T(\theta)$, and wall-clock runtime $W(\theta)$, and compute efficiency metrics as \emph{tokens-per-solved} and \emph{time-per-solved}:
\begin{equation}
\mathrm{Tok/Solved}(\theta)=\frac{T(\theta)}{S(\theta)},\qquad
\mathrm{Time/Solved}(\theta)=\frac{W(\theta)}{S(\theta)},
\label{eq:eff_metrics}
\end{equation}
where $S(\theta)$ denotes the number of problems successfully solved under configuration $\theta$
\footnote{If $S(\theta)=0$, we treat efficiency as $\infty$ for comparison purposes.}.

We retain a compact set of promising configurations by selecting Pareto-efficient points with respect to higher \textit{pass@1}, lower tokens-per-solved, and lower time-per-solved. Accordingly, we define the objective vector as
\begin{equation}
\mathbf{m}(\theta)=\big(p(\theta),\ -\mathrm{Tok/Solved}(\theta),\ -\mathrm{Time/Solved}(\theta)\big).
\label{eq:pareto_metric_vec}
\end{equation}
We say that a configuration $\theta_i$ \emph{dominates} $\theta_j$ (denoted $\theta_i \succ \theta_j$) if it is no worse in all objectives and strictly better in at least one, i.e.,
\begin{equation}
\forall k,\ m_k(\theta_i)\ge m_k(\theta_j)\ \ \wedge\ \ \exists k,\ m_k(\theta_i)> m_k(\theta_j).
\label{eq:pareto_dom}
\end{equation}
Given a set of evaluated configurations $\Theta$, the Pareto frontier is defined as the subset of configurations that are not dominated by any other in $\Theta$:
\begin{equation}
\mathcal{F}(\Theta)=\{\theta\in\Theta\mid \nexists\theta'\in\Theta:\theta'\succ\theta\}.
\label{eq:pareto_frontier}
\end{equation}

In Stage~1, we identify the Pareto-efficient set $\mathcal{F}(\Theta_1)$ and select the Top-$K$ configurations for further refinement. Stage~2 (local refinement) constructs a denser joint grid around each selected candidate, varying $D$ by $\pm1$ and fine-tuning $C$ and $\tau$ within a bounded local range. The resulting configurations are merged, deduplicated, and re-evaluated with additional seeds to improve stability. From the Stage-2 Pareto frontier $\mathcal{F}(\Theta_2)$, the final hyperparameter setting $(D^{*}, C^{*}, \tau^{*})$ is chosen by prioritizing the highest \textit{pass@1}, with ties broken by minimizing tokens-per-solved and then time-per-solved. This two-stage procedure yields a principled, compute-aware selection of depth and exploration--exploitation hyperparameters.

\begin{table*}[!htbp]
\centering
\caption{Stage-1 coarse screening results on CodeContests, summarizing performance across different MCTS depth ($D$), exploration coefficient ($C$), and temperature ($\tau$) configurations.}
\label{tab:rq3_stage1}
\begin{tabular}{
            >{\centering\arraybackslash}p{1.5cm}
            >{\centering\arraybackslash}p{1.5cm}
            >{\centering\arraybackslash}p{1.5cm}
            >{\centering\arraybackslash}p{2.8cm}|
            >{\centering\arraybackslash}p{2.8cm}
            >{\centering\arraybackslash}p{2.8cm}}
\toprule
$D$ & $C$ & $\tau$ & \textit{pass@1} (\%) $\uparrow$ & $Tok/Solved(\theta)$ $\downarrow$ & $Time/Solved(\theta)$ $\downarrow$ \\
\midrule
4 & 0.5 & 0.4 & 35.15 & 125 & 5.6 \\
4 & 1.0 & 0.6 & 40.00 & 118 & 5.2 \\
4 & 1.5 & 0.8 & 40.61 & 115 & 5.1 \\
5 & 0.5 & 0.4 & 36.36 & 122 & 5.5 \\
5 & 1.0 & 0.6 & 41.21 & 112 & 4.9 \\
5 & 1.5 & 0.8 & 41.82 & 110 & 4.8 \\
6 & 0.5 & 0.4 & 35.76 & 135 & 6.2 \\
6 & 1.0 & 0.6 & 39.39 & 125 & 5.8 \\
6 & 1.5 & 0.8 & 40.00 & 122 & 5.6 \\
\bottomrule
\end{tabular}
\end{table*}

\textbf{Result.}
Table~\ref{tab:rq3_stage1} summarizes the \textbf{Stage-1 coarse screening} results on CodeContests (165 problems; averaged over two seeds). Increasing the search depth from $D{=}4$ to $D{=}5$ consistently improves effectiveness under comparable budgets, whereas overly deep trees ($D{=}6$) incur higher cost without proportional gains. The best coarse-grid configuration is $(D{=}5,,C{=}1.5,,\tau{=}0.8)$, achieving a \textit{pass@1} of 41.82\%, compared with 40.00\% for the default baseline (4,1.0,0.6). Efficiency metrics also improve: tokens-per-solved decreases from 118 to 110, and time-per-solved from 5.2 to 4.8, indicating that moderate depth combined with slightly stronger exploration benefits both accuracy and resource utilization.

Table~\ref{tab:rq3_main_results} shows the \textbf{Stage-2 local refinement} outcomes. After refining the neighborhood of the Stage-1 Pareto candidates and re-evaluating with additional seeds, the recommended hyperparameter setting is $(D^{}{=}5$,$C^{}{=}1.4$,$\tau^{}{=}0.7)$, yielding a \textit{pass@1} of \textbf{42.42\%} (70/165). This represents a +2.42 point improvement over the default baseline (40.00\%) while also lowering efficiency cost (tokens-per-solved: 118 $\rightarrow$ 111; time-per-solved: 5.2 $\rightarrow$ 4.9). An optional efficiency-focused Pareto point (5, 1.2, 0.6) achieves 41.82\% with the lowest cost (104 tokens-per-solved; 4.6 time-per-solved), illustrating the expected trade-off between effectiveness and efficiency. Following our selection rule, which prioritizes maximal \textit{pass@1} and breaks ties first by $Tok/Solved$ and then by $Time/Solved$, we adopt $(D^{},C^{},\tau^{})$ as the final hyperparameter setting.

\begin{table*}[!htbp]
\centering
\caption{Recommended MCTS configuration $(D^{},C^{},\tau^{*})$ and comparative results on CodeContests (165 problems).}
\label{tab:rq3_main_results}
\begin{tabular}{l
                >{\centering\arraybackslash}p{0.5cm}
                >{\centering\arraybackslash}p{0.5cm}
                >{\centering\arraybackslash}p{0.5cm}
                |
                >{\centering\arraybackslash}p{2.7cm}
                >{\centering\arraybackslash}p{2.7cm}
                >{\centering\arraybackslash}p{2.7cm}
                }
\toprule
\textbf{Setting} & $D$ & $C$ & $\tau$ & \textit{pass@1} (\%) $\uparrow$ & $Tok/Solved(\theta)$ $\downarrow$ & $Time/Solved(\theta)$ $\downarrow$ \\
\midrule
Default (baseline) $\theta_{\mathrm{base}}$ & 4 & 1.0 & 0.6 & 40.00 & 118 & 5.2 \\
Stage-2 recommended $(D^{*},C^{*},\tau^{*})$ & 5 & 1.4 & 0.7 & 42.42 & 111 & 4.9 \\
Efficiency-first Pareto point (optional) & 5 & 1.2 & 0.6 & 41.82 & 104 & 4.6 \\
\bottomrule
\end{tabular}
\end{table*}

\begin{tcolorbox}[width=1.0\linewidth, title={Answer to RQ3:}]

MCTS depth and exploration–exploitation hyperparameter jointly govern the trade-off between effectiveness and efficiency. Our results show that increasing depth from shallow to moderate levels improves \textit{pass@1} and lowers cost per solved problem, while further increases yield diminishing returns accompanied by higher runtime and token usage. Likewise, moderate exploration, controlled by $C$ and $\tau$, achieves the best overall performance; insufficient exploration leads to stagnation, whereas excessive exploration raises costs without consistent gains in accuracy.
 
\end{tcolorbox}

\subsection{RQ4: Effectiveness Under Realistic Competitive Contests}

\textbf{Approach.} 
To evaluate RQ4 under realistic competitive programming contests, we use two recent regional contests from Codeforces as real-world benchmarks, treating each full contest as a single evaluation instance: (i) the 2025 ICPC Asia Shenyang Regional Contest\footnote{\url{https://codeforces.com/gym/106252}}, 
and (ii) the 2025 National Invitational of CCPC Fujian, the 12th Fujian Collegiate Programming Contest\footnote{\url{https://codeforces.com/gym/105977}}. 
For each problem in a contest, {\tool} generates up to $k$ candidate solutions under a fixed computational budget. We report \textit{pass@1}, \textit{pass@3}, and \textit{pass@5} as the primary metrics, where \textit{pass@$k$} indicates whether at least one of the first $k$ solutions passes all official test cases.

We adopt two complementary evaluation protocols to assess {\tool} under matched contest conditions. First, we select the best-performing baseline from RQ1 and evaluate it on the same contests using identical problem sets, contest duration, computational budget, and execution/evaluation harness as used for {\tool}. Contest-level outcomes, including the number of solved problems and \textit{pass@$k$} (pass@1, pass@3, and pass@5), are compared directly to ensure that observed differences reflect methodological advantages rather than evaluation discrepancies. Second, we evaluate {\tool} using an ICPC-style medal scoring scheme. Specifically, within the fixed contest duration, solving at least $X_{\text{gold}}$ problems corresponds to a gold medal, solving at least $X_{\text{silver}}$ corresponds to a silver medal, and solving at least $X_{\text{bronze}}$ corresponds to a bronze medal, with remaining cases categorized accordingly. For each \textit{pass@$k$} setting, we aggregate the number of solved problems per contest and report the highest medal tier achieved by {\tool}. Medal assignments follow contest-specific award information from the XCPCIO scoreboard platform\footnote{\url{https://board.xcpcio.com/}}, which provides team rankings and corresponding medal tiers.

\textbf{Result.} 
Table~\ref{tab:rq4_contest_results} presents the contest-level outcomes on the two real-world benchmarks. Under strictly matched conditions, {\tool} consistently outperforms the strongest baseline identified in RQ1. On the 2025 ICPC Asia Shenyang Regional contest, {\tool} achieves higher \textit{pass@1/3/5} scores (3/13, 6/13, 7/13 vs.\ 1/13, 4/13, 5/13) and solves more problems across each $k$ setting (Solved@1/3/5: 3/6/7 vs.\ 1/4/5). A similar pattern holds for the 2025 CCPC (Fujian) Invitational, where {\tool} attains superior \textit{pass@1/3/5} (4/13, 5/13, 7/13 vs.\ 2/13, 2/13, 5/13) and higher solved counts (4/5/7 vs.\ 2/2/5). These consistent improvements indicate that {\tool}’s combination of blackboard-driven evidence reuse and MCTS-based global planning enhances its ability to solve contest problems compared with the best baseline configuration under the same realistic conditions.

In addition to \textit{pass@$k$} and solved counts, we interpret performance through an ICPC-style medal-based scoring framework aligned with the XCPCIO scoreboard. Under this human-facing metric, {\tool} achieves at least a \textbf{bronze} tier on the evaluated contests (Medal@1/3/5: --/\textbf{Bronze}/\textbf{Silver} for Shenyang; --/\textbf{Bronze}/\textbf{Silver} for Fujian), demonstrating that its end-to-end contest performance is comparable to lower-tier human teams. Overall, these results show that {\tool} not only surpasses the strongest baseline in direct head-to-head comparisons but also attains practically meaningful performance in full contest-style evaluations.

\begin{tcolorbox}[width=1.0\linewidth, title={Answer to RQ4:}]

On two recent regional contests, {\tool} consistently outperforms the strongest baseline in both \textit{pass@k} and solved-problem counts, achieving higher XCPCIO medal tiers as $k$ increases. This demonstrates that {\tool}’s blackboard-driven evidence accumulation, combined with MCTS-based global planning, enables more effective end-to-end contest performance under realistic conditions.

\end{tcolorbox}

\begin{table*}[!htbp]
\centering
\caption{Results on real-world competitive programming contests, comparing {\tool} with the strongest RQ1 baseline under identical contest settings. Reported metrics include \textit{pass@k}, number of solved problems, and medal tiers based on the XCPCIO scoreboard.}
\label{tab:rq4_contest_results}
\small
\setlength{\tabcolsep}{4.5pt}
\begin{tabular}{l l
                >{\centering\arraybackslash}p{1cm}
                >{\centering\arraybackslash}p{1cm}
                >{\centering\arraybackslash}p{1cm}
                >{\centering\arraybackslash}p{1cm}
                >{\centering\arraybackslash}p{1cm}
                >{\centering\arraybackslash}p{1cm}
                >{\centering\arraybackslash}p{1cm}
                >{\centering\arraybackslash}p{1cm}
                >{\centering\arraybackslash}p{1cm}}
\toprule
\multirow{2}{*}{Contest} & \multirow{2}{*}{Method} 
& \multicolumn{3}{c}{pass@k $\uparrow$} 
& \multicolumn{3}{c}{Solved $\uparrow$} 
& \multicolumn{3}{c}{Medal Tier (XCPCIO)} \\
\cmidrule(lr){3-5}\cmidrule(lr){6-8}\cmidrule(lr){9-11}
& & pass@1 & pass@3 & pass@5 & @1 & @3 & @5 & @1 & @3 & @5 \\
\midrule

\multirow{2}{*}{\begin{tabular}[c]{@{}l@{}}2025 ICPC Asia\\Shenyang Regional\end{tabular}}
&  {\tool} & 3/13 & 6/13 & 7/13 & 3 & 6 & 7 & -- & Bronze & Silver \\
& CodeSim & 1/13 & 4/13 & 5/13 & 1 & 4 & 5 & -- & -- & Bronze \\
\midrule

\multirow{2}{*}{\begin{tabular}[c]{@{}l@{}}2025 CCPC (Fujian)\\Invitational\end{tabular}}
&  {\tool} & 4/13 & 5/13 & 7/13 & 4 & 5 & 7 & -- & Bronze & Silver \\
& CodeSim & 2/13 & 2/13 & 5/13 & 2 & 2 & 5 & -- & -- & Bronze \\
\midrule

\multirow{2}{*}{\begin{tabular}[c]{@{}l@{}}\textbf{Overall}\end{tabular}}
&  {\tool} & 7/26 & 11/26 & 14/26 & 7 & 11 & 14 &  &  &  \\
& CodeSim & 3/26 & 6/26 & 10/26 & 3 & 6 & 10 &  &  &  \\
\bottomrule
\end{tabular}
\end{table*}

\section{Discussions}
\label{sec:analysis}

\subsection{Token Assumption Analysis}

\subsubsection{Agent-level Token Consumption Statistics}

\textbf{Motivation.} We include a per-agent token breakdown to understand \emph{where} computation is spent in a multi-agent search system and whether the observed gains are driven by evidence-producing components such as testing and repair rather than repeated generic generation. This analysis also reveals potential inefficiencies, for example, agents that consume substantial tokens without proportionate improvements, which helps identify opportunities for future optimization.

\textbf{Approach.} To quantify agent-level token consumption statistics across datasets, we run {\tool} on all four datasets and log token usage for every LLM call, tagging each call with the agent that triggered it.  For each dataset, we aggregate prompt and completion tokens per agent over all problem instances, all iterations, and all search branches, and then compute the average per-problem usage for each agent, reported as Average Prompt Tokens and Average Completion Tokens.  We further report each agent's share of the average per-problem token budget by normalizing these averages against the dataset-level per-problem totals.  Throughout, we apply a consistent token accounting protocol across agents and phases, so the breakdown reflects true compute allocation rather than differences in measurement.

\begin{figure*}[t]
  \centering
  \subfloat[Average Prompt Tokens (per--problem).]{
    \includegraphics[width=0.90\textwidth]{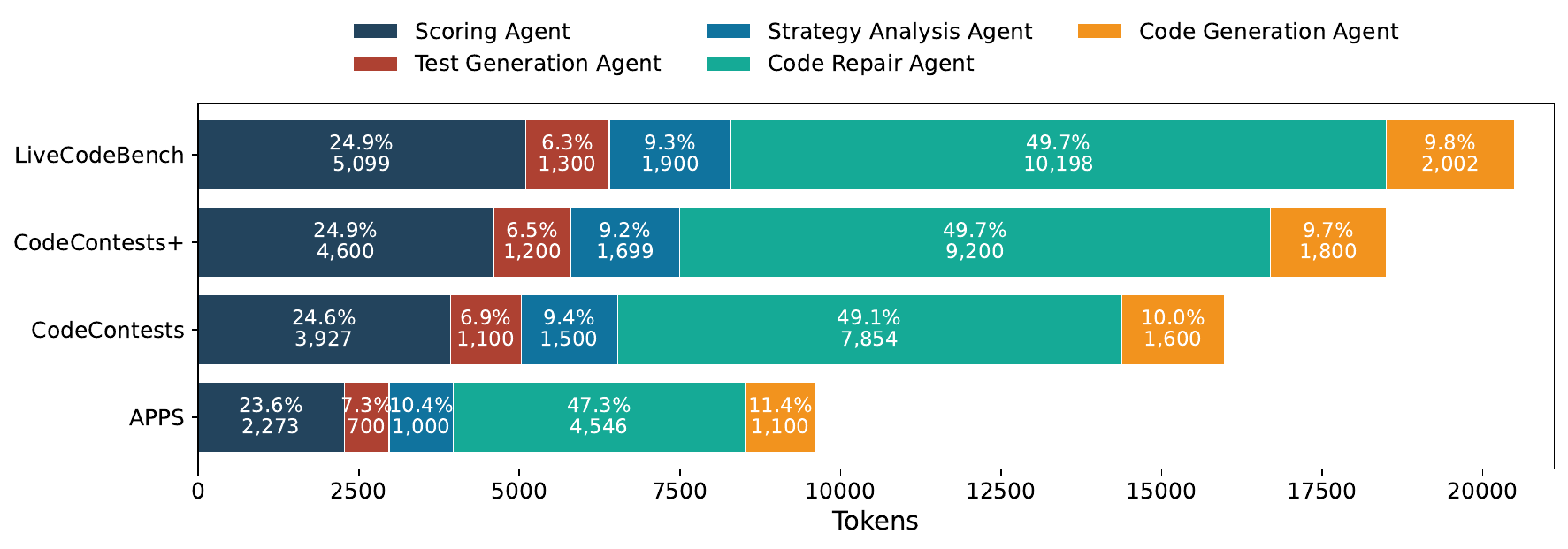}
    \label{fig:agent_tokens_prompt}
  }
  \hfill
  \subfloat[Average Completion tokens (per--problem).]{
    \includegraphics[width=0.90\textwidth]{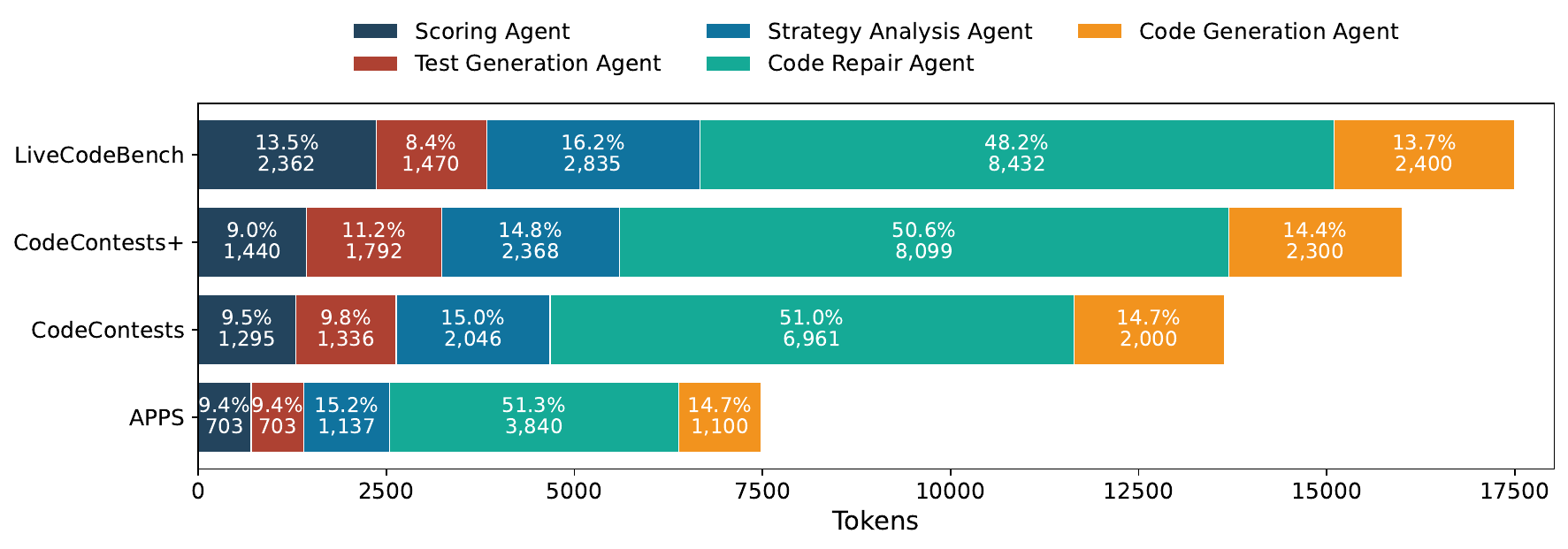}
    \label{fig:agent_tokens_completion}
  }
  \caption{Agent-level token consumption across datasets.}
  \label{fig:agent_token_breakdown}
\end{figure*}

\textbf{Result.} As shown in Figure~\ref{fig:agent_token_breakdown}, prompt-token usage is dominated by the Scoring Agent and the Code Repair Agent across all datasets. On APPS, the Scoring Agent and Code Repair Agent consume 2{,}273 and 4{,}546 prompt tokens per problem, respectively, while Test Generation and Strategy Analysis remain much smaller at 700 and 1{,}000, and Code Generation is 1{,}100. The same pattern holds on harder benchmarks. On LiveCodeBench, Scoring and Code Repair increase to 5{,}099 and 10{,}198 prompt tokens per problem, compared with 1{,}300 for Test Generation, 1{,}900 for Strategy Analysis, and 2{,}002 for Code Generation. This is expected because these two agents are triggered more frequently during search. In particular, scoring is invoked whenever a new draft is produced to obtain execution feedback, and repair becomes the key driver as drafts approach correctness. Moreover, both agents must repeatedly carry the full program text in their inputs, so their prompt tokens grow substantially with the number of invocations and the amount of contextual evidence attached.

For completion tokens, the largest consumers are the Code Generation Agent and the Code Repair Agent, reflecting that both must emit full code artifacts rather than short structured outputs. On APPS, Code Repair and Code Generation average 3{,}840 and 1{,}100 completion tokens per problem, exceeding Scoring, Test Generation, and Strategy Analysis at 703, 703, and 1{,}137. A similar trend appears on CodeContests, where Code Repair and Code Generation reach 6{,}961 and 2{,}000 completion tokens per problem, while the remaining agents stay in the 1{,}295 to 2{,}046 range. On the harder datasets, Code Repair remains the top consumer, reaching 8{,}099 on CodeContests+ and 8{,}432 on LiveCodeBench, with Code Generation at 2{,}300 and 2{,}400. This pattern follows directly from agent roles. Code Generation is responsible for producing code drafts, while Code Repair handles later-stage iterative patching of the draft code and is typically called more times than Code Generation as the search focuses on fixing concrete failures, leading to the highest completion-token usage overall.

At the dataset level, harder benchmarks require more tokens per problem. CodeContests+ and LiveCodeBench show higher average prompt and completion usage than APPS and CodeContests, which is consistent with increased difficulty and stricter evaluation demanding more extensive exploration and more iterations of testing and repair before converging to a passing solution.

\subsubsection{Budget Fairness and Comparability}

\textbf{Motivation.} Because token budget is a primary confounder in agentic and search-based code generation, improvements in pass rates can be misleading if they are achieved by simply spending more tokens. Our goal is to demonstrate that {\tool}'s gains come from better decision making and more effective allocation of compute across strategy, testing, and repair, rather than a brute-force increase in total token usage. Therefore, we explicitly analyze budget fairness and report cost-normalized metrics to verify that accuracy improvements persist when accounting for resource consumption.

\textbf{Approach.} To evaluate cost-aware comparability, we focus on the strongest baseline identified in our benchmark comparison, CodeSim, and compare it against {\tool} under matched settings, including the same model backend, inference configuration, prompts, tooling, and sandbox environment, and evaluation protocol. Beyond reporting pass rates, we compute efficiency metrics that normalize performance by resource usage, primarily tokens-per-solved, defined as total token consumption divided by the number of solved problems. This cost-normalized view allows us to assess whether {\tool} delivers higher accuracy at comparable or lower token cost per successful solution, rather than benefiting from increased token expenditure.

\textbf{Result.} As shown in Table~\ref{tab:tokens_per_solved_gpt4o}, {\tool} consistently achieves higher Pass@1 than CodeSim across all four datasets under GPT-4o, but with slightly higher cost per successful solve. On APPS, {\tool} solves 62/150 problems with 41.30\% Pass@1, compared to 27/150 and 18.00\% for CodeSim, while its Tokens/Solved is 49{,}163 versus 44{,}398. A similar pattern holds on CodeContests, where {\tool} reaches 77/165 solved and 46.67\% Pass@1, outperforming CodeSim at 43/165 and 26.06\%, with Tokens/Solved of 65{,}095 versus 61{,}625. On the more challenging CodeContests+ and LiveCodeBench benchmarks, {\tool} preserves a clear accuracy advantage, solving 30/110 versus 16/110 and 23/110 versus 13/110, but Tokens/Solved remains modestly higher at 144{,}855 versus 127{,}700 and 180{,}729 versus 158{,}214, respectively.

This tradeoff is expected given {\tool}'s design. Compared with CodeSim, {\tool} allocates additional computation to evidence-driven exploration, including repeated testing, counterexample analysis, and repair iterations, which increases total token usage but improves the probability of escaping early incorrect drafts and converging to a correct solution. Because these diagnostics and repair steps are invoked more frequently on harder datasets, the total tokens per success rises substantially on CodeContests+ and LiveCodeBench for both methods, and the gap in Tokens/Solved reflects that {\tool} spends extra budget to sustain higher solve rates under stricter evaluation.

\begin{table*}[t]
\centering
\small
\caption{Cost-aware comparison between {\tool} and CodeSim under GPT-4o. Solved counts are reported as solved/total. Tokens/Solved is computed as Total Tokens divided by Solved and then floored to an integer.}
\begin{tabular}{l l >{\centering\arraybackslash}p{2cm}
                    >{\centering\arraybackslash}p{2cm}
                    >{\centering\arraybackslash}p{2cm}
                    >{\centering\arraybackslash}p{2cm}}
\toprule
Dataset & Method & Solved & Pass@1 (\%) & Total Tokens & Tokens/Solved \\
\midrule
\multirow{2}{*}{APPS (150)} &
{\tool} & 62/150 & 41.30 & 3{,}048{,}127 & 49{,}163 \\
& CodeSim & 27/150 & 18.00 & 1{,}198{,}765 & 44{,}398 \\
\midrule
\multirow{2}{*}{CodeContests (165)} &
{\tool} & 77/165 & 46.67 & 5{,}012{,}345 & 65{,}095 \\
& CodeSim & 43/165 & 26.06 & 2{,}649{,}876 & 61{,}625 \\
\midrule
\multirow{2}{*}{CodeContests+ (110)} &
{\tool} & 30/110 & 27.27 & 4{,}345{,}678 & 144{,}855 \\
& CodeSim & 16/110 & 14.55 & 2{,}043{,}210 & 127{,}700 \\
\midrule
\multirow{2}{*}{LiveCodeBench (110)} &
{\tool} & 23/110 & 20.91 & 4{,}156{,}789 & 180{,}729 \\
& CodeSim & 13/110 & 11.82 & 2{,}056{,}789 & 158{,}214 \\
\bottomrule
\end{tabular}

\label{tab:tokens_per_solved_gpt4o}
\end{table*}

\subsection{Threats to Validity}

\subsubsection{Internal Validity}

\noindent{\textbf{Compute budget and resource allocation.}}
Subtle mismatches in compute budgeting (e.g., per-step token limits, number of expansions, maximum depth, and budget allocation across analysis/generation/testing/repair) can shift the exploration--exploitation balance in search-based systems, making observed gains partially attributable to budgeting choices rather than algorithmic advantages.
To alleviate this threat, we fix the model backend, inference settings, prompt templates, tooling/sandbox environment, and the quick/deep evaluation pipeline, and enforce a consistent overall budget across all methods to maximize comparability. We additionally report runtime statistics to contextualize accuracy improvements with their computational overhead.

\noindent{\textbf{Faithfulness of semantic translation into the blackboard representation.}}
Translating problem statements into a structured blackboard may omit key details or resolve ambiguities incorrectly, which can systematically bias downstream decisions (e.g., test synthesis and repair). If the blackboard is mutable during search, later operations may introduce silent drift and amplify early interpretation errors across branches.
To alleviate this threat, we adopt a fixed and standardized extraction schema with consistent prompting for blackboard construction, and we treat the problem-model components on the blackboard as immutable state so that subsequent operations cannot modify them, preventing silent drift throughout the search.

\noindent{\textbf{Evaluation pipeline effects.}}
Performance is sensitive to the exact evaluation protocol (sandbox execution, timeouts, test ordering, early stopping, and fast screening). Small implementation differences can flip borderline outcomes and, because evaluation feedback guides search, can alter value propagation and expansion priorities.
To alleviate this threat, we run all methods under the same sandbox and an identical evaluation protocol, fix random seeds where applicable, and keep timeout and stopping rules consistent across all configurations. As future work, we plan to expand evaluations to additional platforms and contest settings to further stress-test robustness.

\noindent{\textbf{Data leakage and contamination.}}
LLMs may have been exposed during pretraining or subsequent data collection to benchmarks, near-duplicates, or related solutions, so apparent gains can partially reflect data leakage or memorization rather than true generalization.
To alleviate this threat, we validate across multiple benchmarks and include LiveCodeBench, which is continuously updated and incorporates contamination auditing, thereby reducing the likelihood that reported gains are explained by leakage.

\subsubsection{External Validity}

\noindent{\textbf{Representativeness of tasks and platforms.}}
Contest-style benchmarks emphasize algorithm selection and edge-case handling, which differ from real-world software engineering in requirements, maintainability, dependencies, and long-term evolution; thus, findings may not directly transfer to broader engineering code generation scenarios.
To alleviate this threat, we evaluate on multiple benchmarks spanning different domains and protocols and further include full contest instances to better approximate real competitive settings. We also scope our claims explicitly to contest-style program generation to avoid over-generalization.

\noindent{\textbf{Limitations of contest-level evaluation.}}
Offline contest proxies (such as pass@k and solved-to-medal mappings) cannot fully capture human competition dynamics.
To alleviate this threat, we report both per-problem metrics (pass@1/pass@k) and contest-level aggregates under fixed submission budgets, and interpret medal tiers as a proxy for offline problem-solving capacity under compute constraints rather than a full simulation of end-to-end human contest behavior. As future work, we will extend contest-level evaluations to additional platforms and contests.

\section{Related Work}
\label{sec:related}

Competitive program generation aims to synthesize correct and efficient solutions for algorithmic problems drawn from programming contests. Unlike general-purpose code generation, it requires explicit algorithm selection, multi-step reasoning, strict time--space constraints, and robust edge-case handling. Accordingly, recent benchmarks focus on algorithmic diversity and execution-level correctness. ProBench~\cite{yang2025probench} collects real contest problems with algorithm tags and difficulty annotations, while TACO~\cite{li2023taco} provides large-scale problems labeled with fine-grained topics and skills. CodeContests+~\cite{wang2025codecontests+} improves evaluation fidelity via higher-quality test construction, and COMPASS~\cite{meaden2025compass} broadens evaluation with multi-dimensional metrics covering correctness, efficiency, and code quality.

A growing line of work studies how to improve LLM performance in this setting beyond one-shot prompting, including agentic decomposition, self-refinement, and search-based generation. Representative systems such as MapCoder~\cite{islam2024mapcoder}, CodeSim~\cite{islam2025codesim}, CodeCoR~\cite{pan2025codecor}, and CodeTree~\cite{li2025codetree} leverage iterative planning, retrieval/analysis signals, and structured exploration to navigate the large space of candidate algorithms and implementations. Nevertheless, empirical studies consistently report that even strong LLMs remain brittle on competitive programming benchmarks, with success limited under basic prompting and substantial room for improvement even with agentic scaffolding~\cite{wei2025evaluating,hossain2025llm}. A key reason is that contest problems rarely admit a single obvious generation path: multiple algorithmic strategies may be plausible, failure modes recur across problems, and effective solving often requires coordinating distinct skills such as strategy selection, targeted test synthesis, and evidence-driven repair.

Despite progress, prior approaches exhibit two recurring limitations. First, exploration is often implicitly anchored to a dominant draft or a single reasoning thread, which biases search toward local variations and makes it hard to coordinate heterogeneous capabilities at the right time (e.g., switching from strategy analysis to counterexample-driven repair). Second, intermediate artifacts (constraints extracted from statements, discovered invariants, counterexamples, and patch rationales) are frequently not represented as persistent, reusable state; consequently, knowledge gained along one branch is weakly transferred to others, reducing sample-efficiency in large program spaces where similar mistakes reappear.

We propose {\tool}, a reward-informed, blackboard-driven search framework for competitive program generation. Our key novelty is to treat solving as \emph{stateful exploration} over a shared, structured blackboard that externalizes and persists problem understanding and debugging evidence (e.g., constraints, corner cases, counterexamples, and repair hypotheses), while using an adaptive MCTS-style controller to explicitly plan and allocate computation across competing actions. This design directly mitigates the limitations above: (1) the controller enables principled exploration of alternative algorithmic strategies instead of over-committing to a single draft, and (2) the persistent blackboard promotes cross-branch knowledge transfer so that failures and fixes discovered in one branch can guide subsequent expansions, improving exploration efficiency and reducing repeated errors.
Empirically, {\tool} consistently outperforms strong agentic and search-based baselines across multiple competitive programming benchmarks and LLM backbones, and the gains translate to stronger end-to-end outcomes on full contest instances, including improved pass@k behavior and higher contest-level performance under increasing submission budgets.

\section{Conclusion and Future Work}
\label{sec:conclusion}

We presented {\tool}, an agentic, blackboard-driven MCTS framework for competitive program generation. Our key novelty is to cast solving as stateful global search over heterogeneous actions. MCTS explicitly plans and allocates compute across alternative strategy, testing, and repair decisions, while a shared blackboard persists structured evidence such as constraints, counterexamples, and repair cues for reuse across branches and iterations. This design addresses two common limitations of prior approaches. It avoids over-committing to a single dominant draft, where an early solution attempt anchors subsequent exploration and the search spends most of its budget on local edits around the same algorithmic hypothesis. It also avoids treating intermediate artifacts as disposable by transferring diagnostic knowledge across iterations. Empirically, under matched settings, {\tool} consistently outperforms strong agentic and search-based baselines across benchmarks, and the gains translate to stronger end-to-end contest outcomes, with improved pass rates and higher contest-level performance that becomes more pronounced as the submission budget increases.

In the future, we first want to make the blackboard representation more faithful and robust, including better ambiguity handling in problem statements and mechanisms to revise earlier interpretations when new evidence contradicts them. We second want to strengthen the planning and learning signals used by search, such as explicit efficiency modeling and multi-objective rewards that capture solution quality beyond correctness. Finally, we want to incorporate learning-guided priors that adapt branching and budget allocation based on the estimated value of information, improving both sample-efficiency and robustness for long-horizon program generation.

\section*{CRediT authorship contribution statement}

\textbf{Minnan Wei:} Conceptualization, Methodology, Software, Validation, Data Curation, Writing-Original Draft.
\textbf{Xiang Chen:} Conceptualization, Methodology, Writing -review \& editing, Supervision.
\textbf{Xiaoshuai Niu:} Data curation, Software, Validation.
\textbf{Siyu Chen:} Data curation, Software, Validation.

\section*{Declaration of competing interest}
The authors declare that they have no known competing financial interests or personal relationships that could have appeared to
influence the work reported in this paper.

\section*{Data availability}
Data will be made available on request.

\section*{Acknowledgments}
Minnan Wei and Xiang Chen have contributed equally
to this work and are co-first authors. Xiang Chen is the corresponding author.
This research was partially supported by the National Natural Science Foundation of China (Grant No. 61202006), and the Postgraduate Research \& Practice Innovation Program of Jiangsu Province (Grant No. SJCX25\_2003).

 \bibliography{mylib}
\bibliographystyle{elsarticle}

\vspace{1cm}

\noindent\textbf{Minnan Wei} 
is currently pursuing a Master's degree at the School of Artificial Intelligence and Computer Science, Nantong University. His research interests include competitive program generation and vulnerability detection.

\par
\vspace{1cm}

\noindent\textbf{Xiang Chen} 
received the B.Sc. degree in the School of Management from Xi'an Jiaotong University, China in 2002. Then he received his M.Sc., and Ph.D. degrees in computer software and theory from Nanjing University, China in 2008 and 2011 respectively. He is currently an Associate Professor at the School of Artificial Intelligence and Computer Science, Nantong University. He has authored or co-authored more than 170 papers in refereed journals or conferences, such as IEEE Transactions on Software Engineering, ACM Transactions on Software Engineering and Methodology, IEEE Transactions on Reliability, Empirical Software Engineering, Information and Software Technology, Journal of Systems and Software, Software Testing, Verification and Reliability, Journal of Software: Evolution and Process, Automated Software Engineering, Software - Practice and Experience, Science of Computer Programming, Computer \& Security, Knowledge-based Systems, Engineering Applications of Artificial Intelligence, International Conference on Software Engineering (ICSE), International Conference on the Foundations of Software Engineering (FSE), International Conference Automated Software Engineering (ASE), International Symposium on Software Testing and Analysis (ISSTA), International Conference on Software Maintenance and Evolution (ICSME), International Conference on Program Comprehension (ICPC), International Symposium on Software Reliability Engineering (ISSRE) and International Conference on Software Analysis, Evolution and Reengineering (SANER). His research interests include software engineering, in particular software testing and maintenance, security vulnerability detection and understanding, large language models for software engineering, software repository mining, and empirical software engineering. He received two ACM SIGSOFT distinguished paper awards in ICSE 2021 and ICPC 2023. He is an editorial board member of Information and Software Technology. More information can be found at:
\url{https://xchencs.github.io/index.html}.

\par
\vspace{1cm}

\noindent\textbf{Xiaoshuai Niu} 
is currently pursuing his Bachelor degree at the School of Artificial Intelligence and Computer Science, Nantong University. His research interests include software
repository mining.

\par
\vspace{1cm}

\noindent\textbf{Siyu Chen} 
is currently pursuing a Master's degree at the School of Artificial Intelligence and Computer Science, Nan-tong University. Her research interests include software vulnerability analysis.

\clearpage
\appendix
\section{Prompt Templates}
\label{app:prompts}


\lstset{
  basicstyle=\ttfamily\small,
  breaklines=true,
  breakatwhitespace=true,
  columns=fullflexible,
  keepspaces=true,
  showstringspaces=false,
  frame=single,
  framerule=0.3pt,
  xleftmargin=0.8em,
  xrightmargin=0.8em
}

\subsection{Strategy Analysis Prompt}
\label{app:prompts_strategy}
\begin{lstlisting}
You are a competitive programming strategist.

Given the problem and observed evidence, propose algorithmic strategies.

PROBLEM SUMMARY
{problem_summary}

CONSTRAINTS
{constraints}

EVIDENCE FROM EXECUTION
- recent_statuses: {recent_statuses}
- common_failure_patterns: {failure_patterns}
- representative_counterexamples:
{counterexamples}

OUTPUT FORMAT (STRICT JSON)
Return a JSON object with key "strategies", value is a list of strategies.
Each strategy must contain:
- "id": short snake_case id
- "name": short name
- "applicability_conditions": list of strings
- "complexity_upper_bound": string like "O(n log n)"
- "risk_flags": list of strings
- "minimal_evidence_set": list of strings
- "notes": string
- "bid": { "p": float in [0,1], "c": float in [0,1], "r": float in [0,1] }
Also include "recommended_active_id": one of the ids.

RULES
- Provide 2 to 4 strategies.
- If uncertain, include a baseline safe strategy.
- Use constraints to justify complexity bounds.
Return ONLY JSON.
\end{lstlisting}

\subsection{Code Generation Prompt}
\label{app:prompts_codegen}
\begin{lstlisting}
You are an expert competitive programmer.

TASK
Write a correct and efficient Python 3 solution for the problem below.

PROBLEM STATEMENT
{problem_statement}

INPUT/OUTPUT SPEC
{io_spec}

CONSTRAINTS
{constraints}

REQUIRED INVARIANTS / CORRECTNESS CONDITIONS
{invariants}

EDGE CASE CHECKLIST
{edge_cases}

KNOWN FAILING / TRICKY CASES (from blackboard)
{counterexamples}

RULES
- Output MUST be ONLY valid Python code (no markdown, no explanation).
- The solution MUST read from stdin and write to stdout.
- Be robust to extra spaces/newlines.
- Ensure time complexity fits constraints.
- Prefer simple, standard-library-only code.
- Add minimal comments only where needed for correctness.

Return ONLY the final Python code.
\end{lstlisting}

\subsection{Test Generation Prompt}
\label{app:prompts_testgen}
\begin{lstlisting}
You are a test engineer for competitive programming solutions.

PROBLEM
{problem_statement}

INPUT/OUTPUT SPEC
{io_spec}

CONSTRAINTS
{constraints}

EDGE CASE CHECKLIST
{edge_cases}

KNOWN COUNTEREXAMPLES (inputs that broke solutions)
{counterexamples}

TASK
Propose additional test cases that are likely to reveal bugs:
- extreme values
- boundary conditions
- tricky formatting
- small exhaustive cases if applicable

OUTPUT FORMAT (STRICT JSON)
Return a JSON object with key "tests", value is a list.
Each item:
- "input": string (must end with newline)
- "expected_output": optional string (if you can deduce it confidently; else null)
- "origin": one of ["GENERATED_EXTREME","GENERATED_RANDOM",
  "GENERATED_ENUM","MINIMIZATION_HINT"]
- "rationale": short string

RULES
- Provide 6 to 12 tests.
- At least 3 must be extreme/boundary.
- If expected_output is unknown, set it to null.
Return ONLY JSON.
\end{lstlisting}

\subsection{Scoring Prompt}
\label{apps:prompts_scoring}
\begin{lstlisting}
SCORING_PROMPT_TEMPLATE = """You are a strict evaluator for competitive programming solutions.

PROBLEM
{problem_statement}

INPUT/OUTPUT SPEC
{io_spec}

CONSTRAINTS
{constraints}

CANDIDATE CODE
{current_code}

KNOWN COUNTEREXAMPLES
{counterexamples}

TESTS (STRICT JSON)
{tests_json}

EXECUTION RESULTS (STRICT JSON)
{exec_results_json}

TASK
Summarize the evaluation outcome and extract reusable diagnostics.

OUTPUT FORMAT (STRICT JSON)
Return a JSON object with:
- "status": one of ["PASS","FAIL","RUNTIME_ERROR",
    "TIMEOUT","MEMORY_ERROR"]
- "error_type": one of ["WA","TLE","RE","MLE","FORMAT","UNKNOWN"]
- "reward": float in [0,1]
- "failing_tests": list (up to 6 items), each item includes:
"input", "expected_output", "actual_output", "origin", "diff_summary"
- "patch_cues": list of short repair cues grounded in the failures (up to 6)

RULES
- If all tests pass, set status="PASS", error_type="UNKNOWN", reward=1.0, and failing_tests=[].
- If any test fails, set status="FAIL" unless execution indicates RE/TLE/MLE/FORMAT.
- Keep outputs concise and strictly follow the JSON schema.

Return ONLY JSON.
"""
\end{lstlisting}

\subsection{Code Repair Prompt}
\label{app:prompts_repair}
\begin{lstlisting}
You are a senior engineer fixing a competitive programming solution.

PROBLEM (for reference)
{problem_statement}

CURRENT CODE
{current_code}

FAILURE DIAGNOSTICS
- status: {status}
- error_type: {error_type}
- failing_tests:
{failing_tests}

PATCH PROPOSALS (apply the best subset, respect constraints)
{patch_proposals}

RULES
- Return ONLY valid Python code (no markdown, no explanation).
- Preserve working parts; change minimal lines necessary.
- Ensure the fix addresses the failing tests.
- Do NOT introduce new I/O format changes.
- Keep complexity within constraints.
- If multiple patches conflict, choose the safer one.

Return ONLY the repaired Python code.
\end{lstlisting}

\end{document}